\documentclass[aps,prl,twocolumn, groupedaddress,showpacs,amsmath,floatfix]{revtex4}
 \pdfoutput=1
\usepackage{amssymb}
\usepackage{graphicx}
\usepackage{dcolumn}
\usepackage{subfigure}
\usepackage{wrapfig}
\usepackage{cancel}
\usepackage{color}

\newcommand{\nmlda}{{LDA[nm]}}
\newcommand{\nmgga}{{GGA[nm]}}
\newcommand{\fmldau}{{LDA+U[fm]}}
\newcommand{\fmlda}{{LDA[fm]}}
\newcommand{\fmggau}{{GGA+U[fm]}}
\newcommand{\fmgga}{{GGA[fm]}}
\newcommand{\afmggau}{{GGA+U[afm]}}
\newcommand{\afmgga}{{GGA[afm]}}

\begin{document}

\title{ \bf Determination of effective microscopic models for the
  frustrated antiferromagnets Cs$_2$CuCl$_4$ and Cs$_2$CuBr$_4$ by density functional methods}
\author{Kateryna Foyevtsova}
\affiliation{Institut f\"ur Theoretische Physik, Goethe-Universit\"at
  Frankfurt, 60438 Frankfurt am Main, Germany}
\author{Ingo Opahle}
\affiliation{Institut f\"ur Theoretische Physik, Goethe-Universit\"at
  Frankfurt, 60438 Frankfurt am Main, Germany}
\author{Yu-Zhong Zhang}
\affiliation{Institut f\"ur Theoretische Physik, Goethe-Universit\"at
  Frankfurt, 60438 Frankfurt am Main, Germany}
\author{Harald O. Jeschke}
\affiliation{Institut f\"ur Theoretische Physik, Goethe-Universit\"at
  Frankfurt, 60438 Frankfurt am Main, Germany}
\author{Roser Valent\'{\i}}
\affiliation{Institut f\"ur Theoretische Physik, Goethe-Universit\"at
  Frankfurt, 60438 Frankfurt am Main, Germany}

\date{\today}

\begin{abstract}
We investigate the electronic and magnetic properties of the
frustrated triangular-lattice antiferromagnets Cs$_2$CuCl$_4$ and
Cs$_2$CuBr$_4$ in the framework of density functional theory.
Analysis of the exchange couplings $J$ and $J'$ using the available
X-ray structural data corroborates the values obtained
from experimental results for Cs$_2$CuBr$_4$ but not for
Cs$_2$CuCl$_4$.  In order to understand this discrepancy, we perform a
detailed study of the effect of structural optimization on the
exchange couplings of Cs$_2$CuCl$_4$ employing different
exchange-correlation functionals.
We find that the exchange couplings depend on rather subtle details
of the structural optimization and that only when the insulating state 
(mediated through spin polarization) is present in the structural
optimization, we do have  good agreement between the calculated and
 the experimentally determined exchange couplings.
 Finally, we discuss the effect of interlayer couplings as
well as longer-ranged couplings in both systems.
\end{abstract}


\maketitle

\section{I. Introduction}
\begin{figure}
\begin{center}
\subfigure {\includegraphics[width=0.26\textwidth]{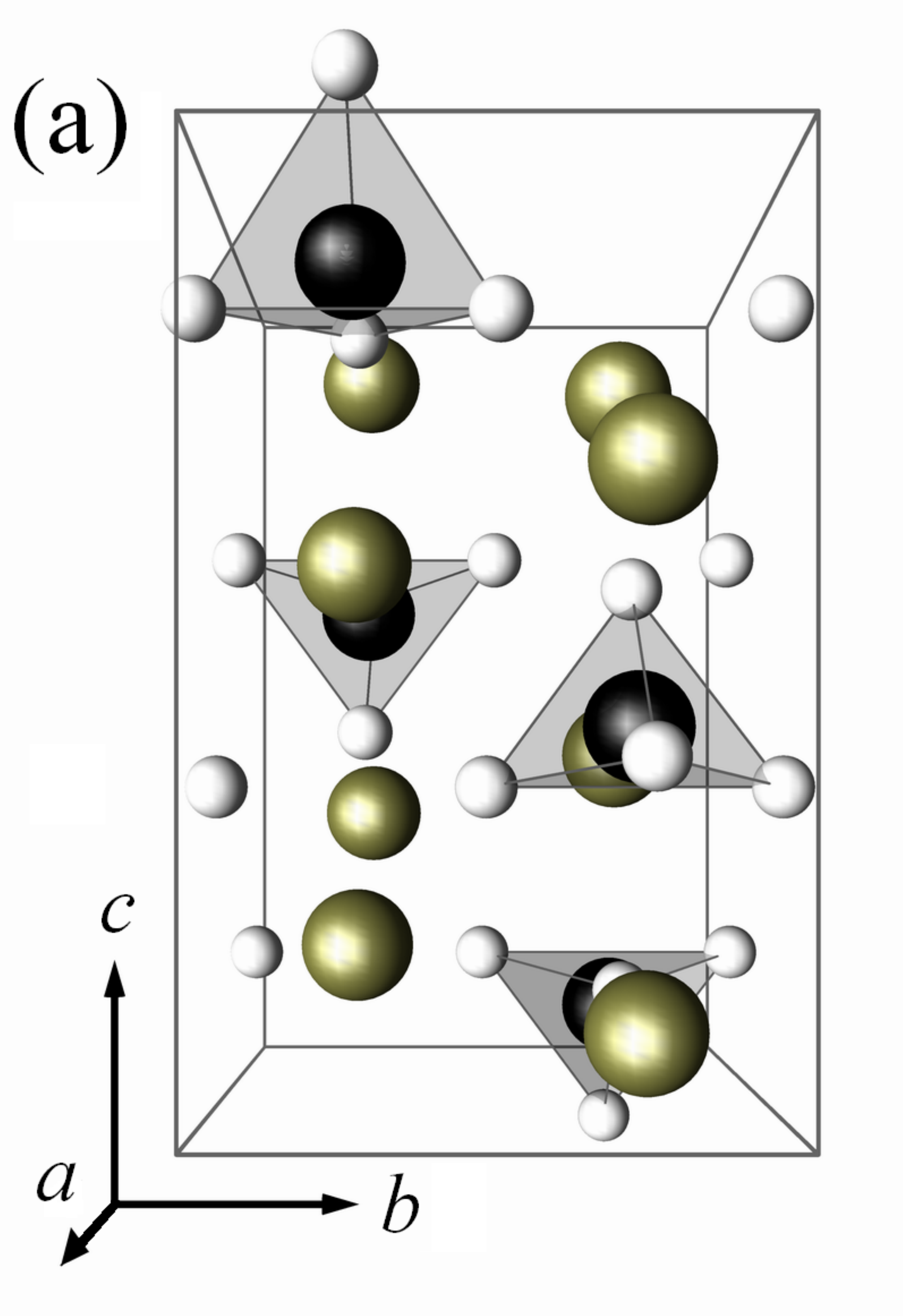}}%
\subfigure {\includegraphics[width=0.26\textwidth]{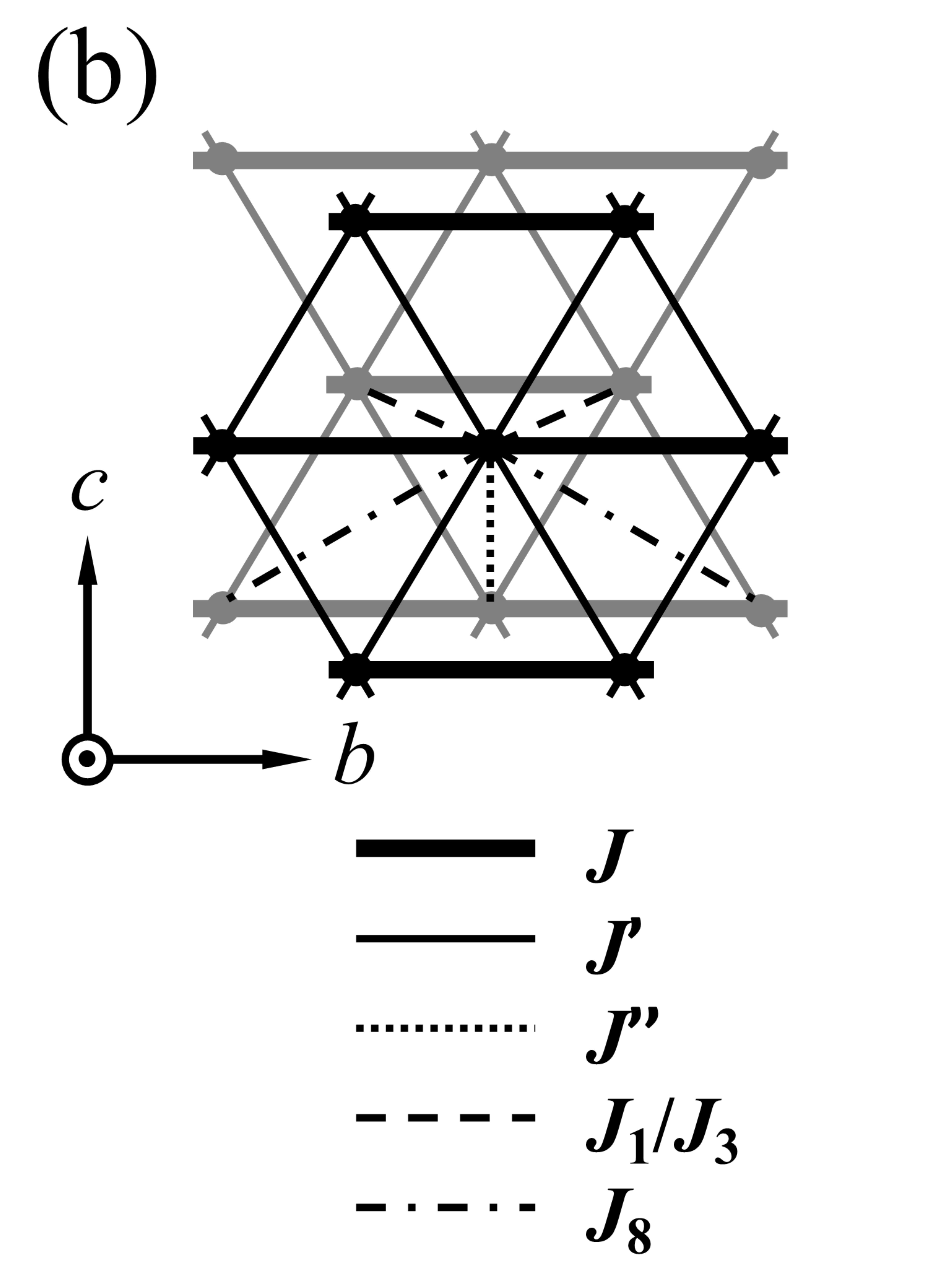}}
\subfigure {\includegraphics[trim = 0mm 8mm 0mm 8mm, clip, width=0.50\textwidth]{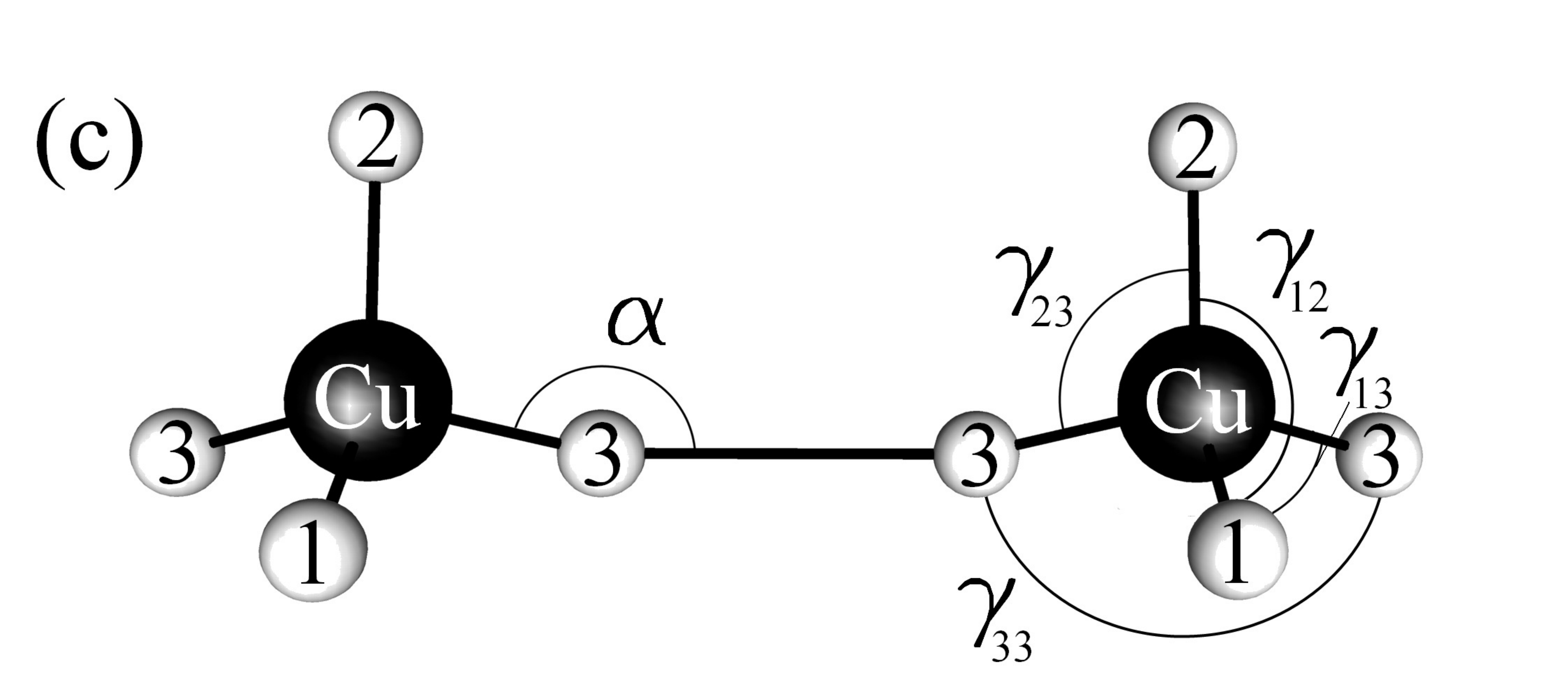}}
\subfigure {\includegraphics[width=0.36\textwidth]{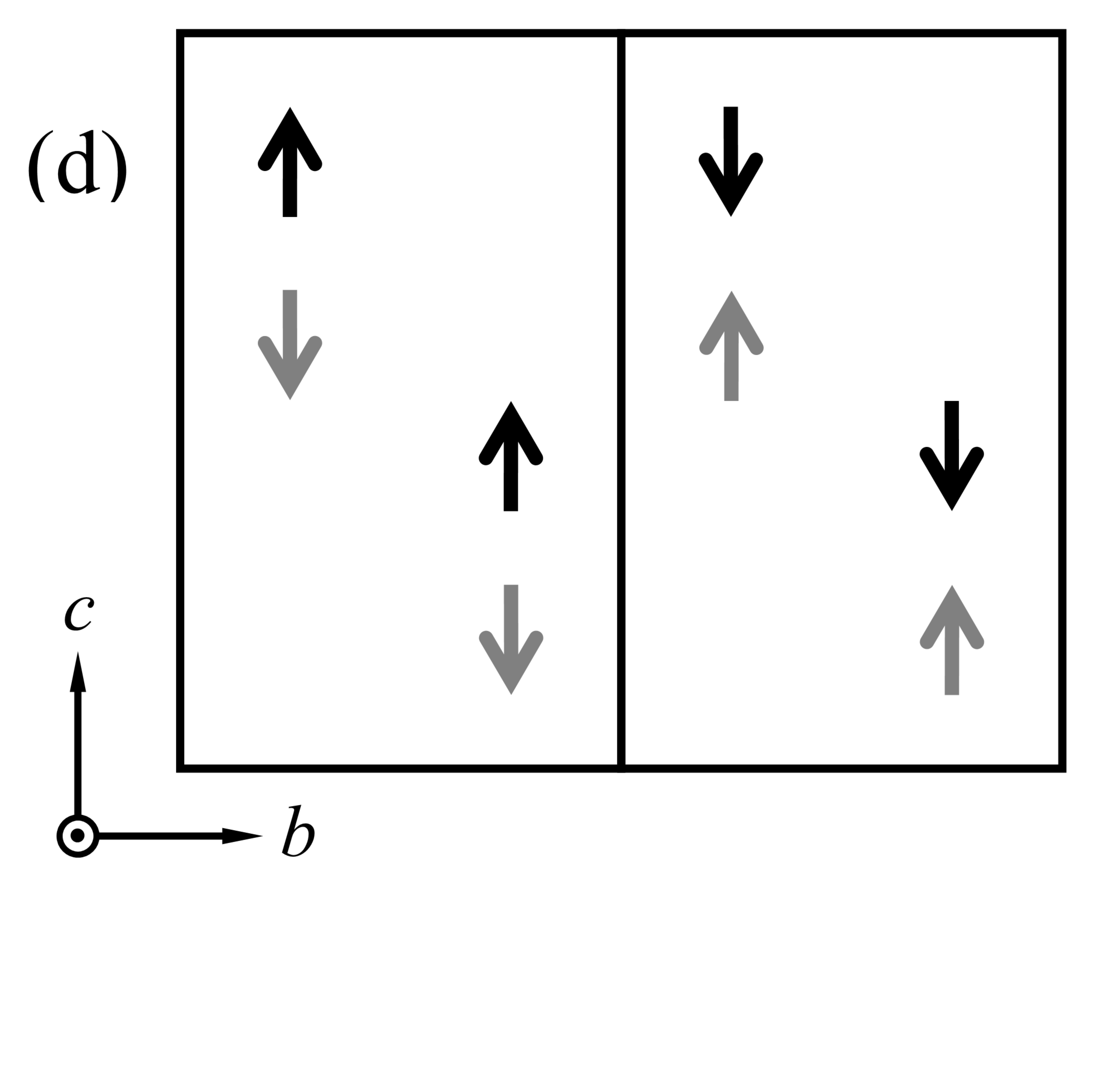}}
\caption{(Color online) (a) The Cs$_2$CuCl$_4$ and Cs$_2$CuBr$_4$ unit cell where the 
  Cu$X_4$ tetrahedra ($X=$ Cl, Br) are highlighted.  (b) Schematic
  triangular lattice of  Cu spins. The
  gray lattice is below the black lattice.  (c) Neighboring Cu$X_4$
  tetrahedra. Labels 1, 2 and 3 of the $X$ atoms denote the three
  inequivalent $X$ atoms: $X$(1), $X$(2) and $X$(3). As indicated,
  angles $\gamma_{12}$, $\gamma_{13}$ correspond to the angles
  $X$(1)-Cu-$X$(2), $X$(1)-Cu-$X$(3), {\it etc.}  (d)  Cu spin
  configuration of a $1\times2\times1$ supercell of Cs$_2$CuCl$_4$
  adopted for the structural optimization within the {\afmggau}
  scheme.}\label{crystal}
\end{center}
\end{figure}

For almost two decades the frustrated antiferromagnets Cs$_2$CuCl$_4$
and Cs$_2$CuBr$_4$ have been considered as experimental realizations
of a frustrated triangular lattice \cite{Coldea96,Tanaka02}. Both
systems crystallize in the space group {\it Pnma}
\cite{Bailleul91,Morosin60} [see Fig.~\ref{crystal}~(a)] and are
characterized by a layered arrangement of Cu$^{2+}$ ions in a
triangular pattern parallel to the $bc$ plane. The two-dimensional
character of magnetic interactions between the spin-$\frac{1}{2}$
Cu$^{2+}$ ions was confirmed by neutron scattering and susceptibility
measurements in both systems
\cite{Coldea03,Tokiwa06,Ono03,Ono05,Tsujii07} and was successfully
modeled \cite{Zheng05,Coldea02} by a two-dimensional Heisenberg
Hamiltonian containing an anisotropic interaction term of the
Dzyaloshinskii-Moriya type \cite{Dzyaloshinskii58,Moriya60}.

In spite of their structural similarity, Cs$_2$CuCl$_4$ and
Cs$_2$CuBr$_4$ have rather different magnetic behavior. While in
Cs$_2$CuBr$_4$ magnetic excitations are localized and the field
dependent magnetization exhibits two well-defined plateaux
\cite{Ono05}, Cs$_2$CuCl$_4$ shows fractional spin excitations and
spin liquid behavior over a broad temperature range, as revealed in
inelastic neutron scattering experiments \cite{Coldea03}. The
dissimilar behavior between Cs$_2$CuCl$_4$ and Cs$_2$CuBr$_4$ has
often been attributed to their unequal degree of frustration,
determined as the ratio $J'/J$ between the inter-chain exchange
coupling $J'$ and the dominant intra-chain exchange coupling $J$ in
the underlying triangular lattice [see Figs. \ref{crystal}~(b) and
  (c)].

A ratio of $J'/J=0.74$~\cite{Ono05_2} has been suggested for
Cs$_2$CuBr$_4$ by comparing the ordering vector of a helical
incommensurate structure observed in neutron elastic scattering
experiments with the one obtained from inverse temperature series
expansions for a spin-$\frac{1}{2}$ Heisenberg model on an anisotropic
triangular lattice~\cite{Weihong99}.  In contrast, a ratio of
$J'/J=0.34$~\cite{Coldea02} was derived for Cs$_2$CuCl$_4$ from
comparison of spin-wave calculations for the spin-$\frac{1}{2}$
Heisenberg model with the magnetic excitation spectrum observed in
neutron scattering experiments in the presence of an external magnetic
field far above saturation. These observations indicate that
Cs$_2$CuCl$_4$ is less frustrated and more one-dimensional than
Cs$_2$CuBr$_4$.

While a large amount of work has been devoted to the description of
the fractional quantum states and to understanding the phase
transitions of the two-dimensional frustrated spin model with spatial
anisotropy~\cite{Zhou03,Yunoki04,Aliceae05,Veillette05,Isakov05,Zheng06,Dalidovich06,Weng06,Yunoki06,Fjarestad06,Balents07,Starykh07,Alicea07,Alicea09,Fortune09,Kohno09,Heidarian09,Xu09,Bishop09},
a detailed comparative analysis of the electronic, magnetic and
structural properties of these systems as well as a deep understanding
of the origin of the different behavior is still missing.

In this work, we present an extensive density functional theory (DFT)
study of the microscopic properties of Cs$_2$CuCl$_4$ and
Cs$_2$CuBr$_4$ and compare our results with experimental data.  We
consider different exchange-correlation functionals in order to also
investigate the dependence of the electronic, magnetic and structural
properties on these choices.  Our study of the performance of
different exchange-correlation functionals is motivated by the fact
that physical properties of recently discovered high temperature
Fe-based superconductors are extremely sensitive to the details of DFT
calculations~\cite{Singh08,Mazin08,Han09,Yin09,Opahle09,Zhang09,Zhang10}.
Also, DFT studies of a recently topical layered Mott insulator {TiOCl}
~\cite{Pisani05,Pisani07,Zhang08PRB,Mastrogiuseppe09,Zhang08PRL,Canosa09,Prodi09,Zhang09unpublished,Sing09,ZhangJPcm09}
reveals that it is essential to use a suitable exchange-correlation
functional in DFT calculations to describe correctly the behavior of
this system.

Out of the electronic structure calculations we derive a tight-binding
(TB) Hamiltonian and estimate the Heisenberg exchange coupling
constants between Cu ions from total energy calculations.  Our DFT
derived effective models incorporate a larger number of interacting Cu
neighbors compared to the models used for the experimental data
analysis. We show that some of these terms are crucial for
understanding the behavior of Cs$_2$CuCl$_4$ and Cs$_2$CuBr$_4$.

\section{II. Structural and electronic properties}

\subsection{ (a) Structural optimization schemes}
In this paper, we consider the experimentally measured structures of
Cs$_2$CuCl$_4$ and Cs$_2$CuBr$_4$ as well as internal coordinates
obtained from the structural optimization with various
exchange-correlation functionals.  The experimentally measured unit
cell parameters of Cs$_2$CuCl$_4$ are taken from
Ref.~\onlinecite{Bailleul91} and those of Cs$_2$CuBr$_4$ from
Ref.~\onlinecite{Morosin60}. Both structures were measured at room
temperature.

Crystal structure optimization of Cs$_2$CuCl$_4$ and Cs$_2$CuBr$_4$ is
required since calculations with the experimentally determined
structural parameters find quite large interatomic forces, which
indicates that the structures are not in equilibrium within DFT.  This
holds for the calculations with all the considered
exchange-correlation functionals listed below.  In all cases, the
interatomic forces are larger in Cs$_2$CuCl$_4$ compared to
Cs$_2$CuBr$_4$.

Based on our structural optimizations, we will discuss in detail the
electronic properties of both compounds.  For the structure
optimization we apply the following schemes with different
approximations to the exchange-correlation functional within DFT and
different magnetic configurations: (1) the spin-independent local
density approximation ({\nmlda}) \cite{Perdew92}; (2) the
spin-independent generalized gradient approximation ({\nmgga})
\cite{Perdew96}; (3) the spin-dependent GGA with a ferromagnetic Cu
spin configuration ({\fmgga}); (4) the spin-dependent
GGA+U~\cite{Chap03} with a ferromagnetic Cu spin configuration
({\fmggau}). For the {\fmggau} calculations we considered the around
mean field (AMF) version \cite{Czyzyk94} with values of $U$ and
$J_\text{H}$ for the Cu ions of 6~eV and 1~eV respectively.  The
lattice constants, which are assumed to be well determined from
experiments, were kept fixed for the structure relaxations while the
optimization of the relative atomic positions was constrained by the
symmetry of the $Pnma$ space group.

For Cs$_2$CuCl$_4$, we also considered two optimization schemes with
an antiferromagnetic Cu spin configuration: {\afmgga} and {\afmggau}
($U=6$~eV, $J_\text{H}=1$~eV). The spin arrangement in this
antiferromagnetic configuration is shown in Fig.~\ref{crystal} (d),
where the Cs$_2$CuCl$_4$ unit cell was doubled in the $b$ direction
along the Cu chains. In order to produce such an arrangement, the
symmetry of the supercell was lowered to the space group $P21/c$, with
two inequivalent Cu atoms. Our choice of this particular
antiferromagnetic configuration is due to its resemblance to the
experimentally observed 120$^{\circ}$ ground state configuration
\cite{Coldea01}.  The considered antiferromagnetic configuration is
collinear, which is beneficial in terms of computational effort, and
fulfills the requirement that the strongest couplings $J$ are
satisfied and the second strongest couplings $J'$ are partially
satisfied.  The optimization of the relative atomic positions with the
two antiferromagnetic schemes was constrained by the symmetry of the
$P21/c$ space group.

Additionally, for both systems we completed the LDA series of
structural optimizations with {\fmlda} and {\fmldau}
optimizations. However, due to the analogous behavior of the
structural properties of Cs$_2$CuCl$_4$ and Cs$_2$CuBr$_4$ observed
within this series with those within the GGA series, the detailed
analysis of the LDA series will be omitted here.

The DFT structural optimizations were performed with the
full-potential local-orbital (FPLO) code \cite{Koepernik99,fplo} in
the scalar relativistic approximation with up to 512 {\bf k}-points in
the full Brillouin zone.

\subsection{ (b) Band gap}

\begin{figure}[h]
\begin{center}
\includegraphics[width=0.47\textwidth]{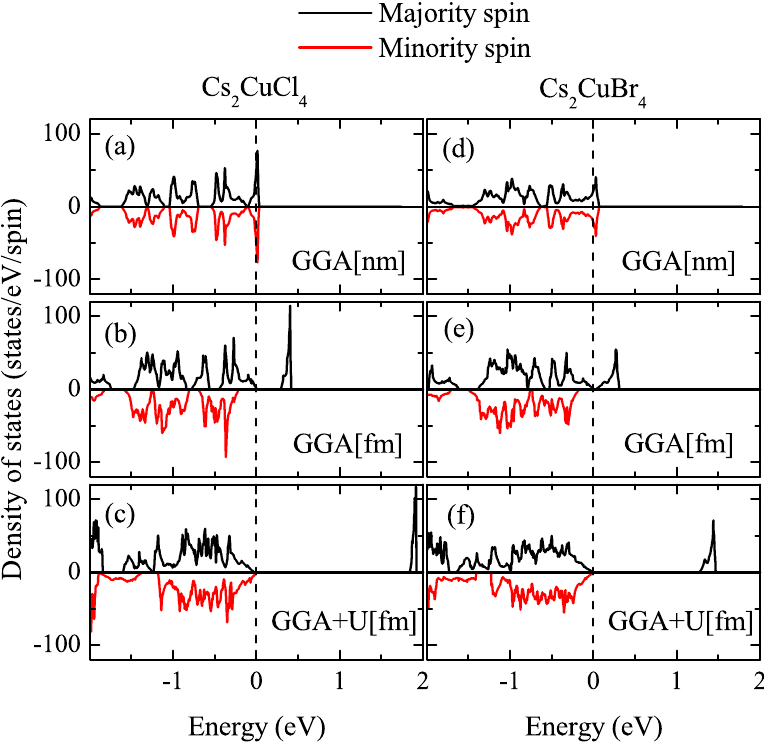}
\caption{(Color online) Total densities of states for majority and minority electron
  spin for (a) Cs$_2$CuCl$_4$, calculated with {\nmgga}, (b)
  Cs$_2$CuCl$_4$, calculated with {\fmgga}, (c) Cs$_2$CuCl$_4$,
  calculated with {\fmggau}, (d) Cs$_2$CuBr$_4$, calculated with
  {\nmgga}, (e) Cs$_2$CuBr$_4$, calculated with {\fmgga}, and (f)
  Cs$_2$CuBr$_4$, calculated with {\fmggau}.  For an easier comparison,
  the two DOS's are plotted with opposite signs.  The calculations are
  performed with the LAPW code Wien2k [see Section II (d)].  In the
  case of {\fmggau}, the around mean field scheme is employed, with
  $U=6$~eV and $J_{\rm H}=1$~eV.  The Fermi level is set to zero.}
\label{gap}
\end{center}
\end{figure}

Before presenting the structure optimizations with the different
exchange-correlation functionals, we shall consider the experimental
crystal structure and discuss the electronic properties obtained with
the different exchange-correlation functionals.  Calculations with the
{\nmgga}, {\fmgga} or {\fmggau} exchange-correlation potentials result in
the electronic structure being either gapless ({\nmgga}) or gapped
({\fmgga} and {\fmggau}), as shown in Fig.~\ref{gap}.  Allowing for spin
polarization opens a gap in both Cs$_2$CuCl$_4$ and Cs$_2$CuBr$_4$
[though in the latter compound the {\fmgga} gap is rather small ($\sim
  0.03$~eV)].  Upon introducing the onsite Coulomb repulsion within
the {\fmggau}, the gaps in both systems increase considerably
[Fig.~\ref{gap} (c) and (f)].

 Spin-dependent exchange-correlation functionals provide a better
 description of Cs$_2$CuCl$_4$ and Cs$_2$CuBr$_4$ electronic
 properties as these functionals correctly reproduce the
 experimentally observed insulating ground state of the
 compounds. Moreover, as it will be shown in the next sections, it
 turns out that the presence of a band gap is important for the
 accurate determination of the equilibrium crystal structures.

\subsection{ (c) Structural analysis}
\begin{figure*}
\begin{center}
\subfigure {\includegraphics[width=1.0\textwidth]{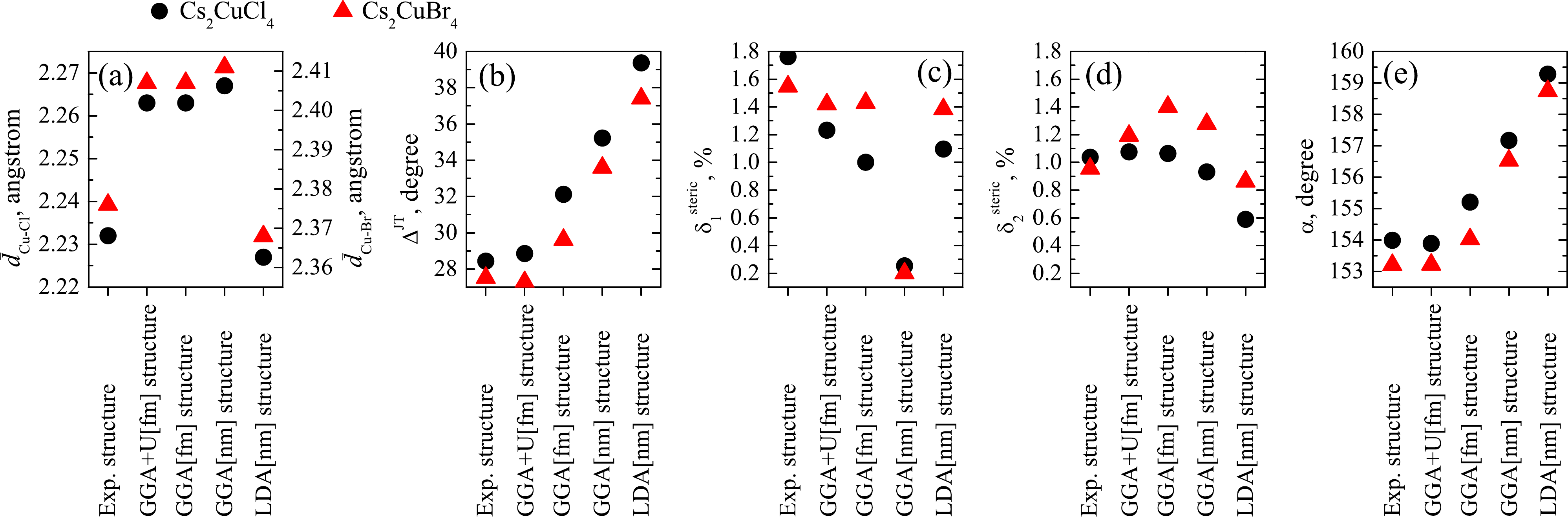}}
\caption{(Color online) (a) The average Cu-$X$ ($X= \text{Cl}$, Br) bond distance
  $\bar{d}_{{\rm Cu}-X}$, (b) Jahn-Teller deviation $\Delta^{\rm JT}$,
  (c) steric pressure deviation $\delta^{\rm steric}_1$, (d) steric
  pressure deviation $\delta^{\rm steric}_2$ and (e) Cu-$X$-$X$ angle
  $\alpha$ in the $J$ superexchange bridge in the experimental as well
  as relaxed Cs$_2$CuCl$_4$ and Cs$_2$CuBr$_4$ crystal structures.
}\label{str_par}
\end{center}
\end{figure*}
We proceed now with the structural optimization.  The crystal
structures of Cs$_2$CuCl$_4$ and Cs$_2$CuBr$_4$ resulting after
optimization with each of the schemes described above are presented in
Appendix A. In our analysis of these structures we focus on geometry
variations of the Cu$X_4$ ($X= \text{Cl}$, Br) tetrahedron, which
determine the strength of important exchange couplings.

In Cs$_2$CuCl$_4$ and Cs$_2$CuBr$_4$, the Cu$X_4$ tetrahedron is
distorted due to the Jahn-Teller effect and also due to the steric
pressure originating from Cs$^+$ ions~\cite{Morosin60}.
The Jahn-Teller effect results in a squeezing of the tetrahedron such
that the $X$-Cu-$X$ bond angles $\gamma_{12}$ and $\gamma_{33}$
increase and the $X$-Cu-$X$ bond angles $\gamma_{13}$ and
$\gamma_{23}$ decrease [Fig.~\ref{crystal} (c)]. The steric pressure
by Cs(2) on $X$(3) causes additional symmetry lowering by increasing
$\gamma_{13}$ and decreasing $\gamma_{33}$. In order to compare the
strengths of the two types of distortions in different structures, one
can define the Jahn-Teller deviation $\Delta^{\rm JT}$ as the
difference between averages $\frac{1}{2}(\gamma_{12}+\gamma_{33})$ and
$\frac{1}{2}(\gamma_{13}+\gamma_{23})$,
\begin{equation}
\Delta^{\rm JT} = \left| \frac{1}{2}(\gamma_{12}+\gamma_{33}) - \frac{1}{2}(\gamma_{13}+\gamma_{23})\right|,
\end{equation}
and the steric pressure deviations $\delta^{\rm steric}_1$ and $\delta^{\rm steric}_2$ as
\begin{equation}
\delta^{\rm steric}_1=\frac{\frac{1}{2}|\gamma_{12}-\gamma_{33}|}{\frac{1}{2}(\gamma_{12}+\gamma_{33})}
\quad \mbox{and} \quad \delta^{\rm steric}_2=\frac{\frac{1}{2}|\gamma_{13}-\gamma_{23}|}{\frac{1}{2}(\gamma_{13}+\gamma_{23})}.
\end{equation}

As can be seen in Appendix A, the Cs$_2$CuCl$_4$ structures relaxed
with {\fmgga} and {\afmgga} are very close as well as the structures
relaxed with {\fmggau} and {\afmggau}. The tetrahedron parameters in
these structures are also similarly close. This demonstrates that,
within the spin-dependent GGA and GGA+U, interatomic forces in
Cs$_2$CuCl$_4$ are very weakly dependent on the actual Cu spin
configuration and in the following we will consider only ferromagnetic
spin-dependent relaxation schemes for Cs$_2$CuCl$_4$ and
Cs$_2$CuBr$_4$. This
result is a consequence of the rather small energy scale of magnetic
interactions between Cu spins, which is much smaller than the difference 
between total energies usually involved in structural relaxations.

In Tables~\ref{tetra_Cl} and \ref{tetra_Br}, we present the
tetrahedron parameters defined above for the experimental and relaxed
crystal structures of Cs$_2$CuCl$_4$ and Cs$_2$CuBr$_4$, respectively.
Some of them are additionally shown in Fig.~\ref{str_par} in order to
facilitate the comparison of the different crystal structures.  In
both compounds, the crystal structures of the GGA relaxation series
({\fmggau}, {\fmgga} and {\nmgga}) are featured by a continuous variation
of the $X$-Cu-$X$ angles.
In terms of these angles, the relaxed crystal structures closest to
the experimental structures are the {\fmggau} structures for both
Cs$_2$CuCl$_4$ and Cs$_2$CuBr$_4$.  However, in terms of Cu-$X$ bond
distances, represented here as the averaged distance $\bar{d}_{{\rm
    Cu}-X}$, the GGA crystal structures are further away from the
experimental structures than the LDA structures [see
  Fig.~\ref{str_par}~(a)].  In the LDA structures (including the
{\fmlda} and {\fmldau} ones, which are not listed in the tables),
$\bar{d}_{{\rm Cu}-X}$ are smaller than the corresponding parameters
of the experimental structures, but the difference is less compared to
the GGA series.  Thus, the GGA relaxation tends to increase the bond
distances in the Cu$X_4$ tetrahedron while the LDA relaxation
decreases them.

In the GGA series, the Jahn-Teller-like distortion of the tetrahedron,
which we evaluate through $\Delta^{\rm JT}$, is strongest in the
{\nmgga} crystal structures of both compounds and decreases
monotonically as the Cu magnetic moment in the spin-resolved
calculations increases ({\fmgga}, {\fmggau}) [see
  Fig.~\ref{str_par}~(b)].  The Cu magnetic moment in Cs$_2$CuCl$_4$
calculated with either {\fmgga} or {\afmgga} schemes is 0.50$\mu_B$ and
increases to 0.78$\mu_B$ when the onsite Coulomb interaction is
switched on within the GGA+U functional (Table~\ref{tetra_Cl}).  In
Cs$_2$CuBr$_4$, the corresponding values of the Cu magnetic moment are
0.42$\mu_B$ and 0.73$\mu_B$~\cite{remark} (Table~\ref{tetra_Br}).  In
the {\nmlda} structures, $\Delta^{\rm JT}$ takes the largest value.  The
distortion due to steric pressure, which is characterized by
$\delta^{\rm steric}_1$ and $\delta^{\rm steric}_2$
[Figs.~\ref{str_par}~(c) and (d)], does not appear to follow any
general rule.
\begin{table}
\caption{Tetrahedron parameters for the Cs$_2$CuCl$_4$ structures and
  corresponding values of the Cu magnetic moment during structural
  relaxation. The angles are given in degrees, the deviations
  $\delta^{\rm steric}_1$ and $\delta^{\rm steric}_2$ in percent and
  the averaged Cu-Cl distance $\bar{d}_{\rm Cu-Cl}$ in
  {\AA}ngstr{\"o}m.}\label{tetra_Cl}
\begin{tabular}{lccccc}
\hline\hline
& exp & {\fmggau} & {\fmgga} & {\nmgga} & {\nmlda} \\
\hline
$\gamma_{12}$ &131.33&130.97&133.05&133.66&135.56 \\
$\gamma_{13}$ &101.67&101.60&100.67&\;\;99.69&\;\;98.26 \\
$\gamma_{23}$ &\;\;99.58&\;\;99.43&\;\;98.55&\;\;97.86&\;\;97.11 \\
$\gamma_{33}$ &126.79&127.78&130.41&134.34&138.56 \\
$\Delta^{\rm JT}$ &\;\;28.44&\;\;28.86&\;\;32.12&\;\;35.22&\;\;39.38 \\
$\alpha$ &153.99&153.89&155.21&157.17&159.28 \\
&&&& \\
$\delta^{\rm steric}_1$ &1.76&1.23&1.00&0.25&1.10 \\
$\delta^{\rm steric}_2$ &1.04&1.08&1.06&0.93&0.59 \\
&&&& \\
$\bar{d}_{\rm Cu-Cl}$ &2.232&2.263&2.263&2.267&2.227 \\
&&&& \\
$\mu$ & - & 0.78$\mu_B$ & 0.50$\mu_B$ & 0 & 0 \\
\hline\hline
\end{tabular}
\end{table}
\begin{table}
\caption{Tetrahedron parameters for the Cs$_2$CuBr$_4$ structures and
  corresponding values of the Cu magnetic moment during structural
  relaxation. The angles are given in degrees, the deviations
  $\delta^{\rm steric}_1$ and $\delta^{\rm steric}_2$ in percent and
  the averaged Cu-Br distance $\bar{d}_{\rm Cu-Br}$ in
  {\AA}ngstr{\"o}m.}\label{tetra_Br}
\begin{tabular}{lccccc}
\hline\hline
& exp & {\fmggau} & {\fmgga} & {\nmgga} & {\nmlda} \\
\hline
$\gamma_{12}$ &130.40&130.06&131.77&132.53& 133.72 \\
$\gamma_{13}$ &102.16&102.16&101.70&100.48& \;\;99.03\\
$\gamma_{23}$ &\;\;99.93&\;\;99.75&\;\;98.89&\;\;97.94&\;\;97.34 \\
$\gamma_{33}$ &126.42&126.42&128.05&133.07& 137.48\\
$\Delta^{\rm JT}$ &\;\;27.52&\;\;27.28&\;\;29.62&\;\;33.59& \;\;37.41\\
$\alpha$ &153.21&153.22&154.03&156.53&158.74 \\
&&&& \\
$\delta^{\rm steric}_1$ &1.55&1.42&1.43&0.20& 1.38\\
$\delta^{\rm steric}_2$ &0.96&1.19&1.40&1.28& 0.86\\
&&&& \\
$\bar{d}_{\rm Cu-Br}$ &2.376&2.407&2.407&2.411& 2.368\\
&&&& \\
$\mu$ & - & 0.73$\mu_B$ & 0.42$\mu_B$ & 0 & 0 \\
\hline\hline
\end{tabular}
\end{table}

For the further discussion of the electronic structure it is useful to
consider an additional structural parameter, namely, the angle in the
$J$ superexchange bridge Cu-$X$-$X$-Cu, shown in Fig.~\ref{crystal}
(c) as $\alpha$.  This angle is closely related to $\gamma_{33}$ and
behaves analogously in the various structures [see
  Fig.~\ref{str_par}~(e) and Tab.~\ref{tetra_Cl}].  The large
variation of the superexchange coupling $J$ in different
Cs$_2$CuCl$_4$ relaxed crystal structures is mainly attributed to the
variation of $\alpha$, as will be discussed in Section III.

\subsection{ (d) Electronic structure}
In this section, we present a detailed analysis of electronic
properties of Cs$_2$CuCl$_4$ and Cs$_2$CuBr$_4$ for the experimental
and relaxed crystal structures introduced in the previous section.  We
show the results obtained with the spin-independent GGA
exchange-correlation functional as these are also used for the
tight-binding parametrization.  Calculations were performed with both
the FPLO code as well as the linearized augmented plane wave (LAPW)
scheme, as implemented in the Wien2k code \cite{Wien2k}.  The
calculations with both codes are in good agreement and the data
presented here are obtained with the LAPW code.

\begin{figure}[h]
\begin{center}
\includegraphics[width=0.45\textwidth]{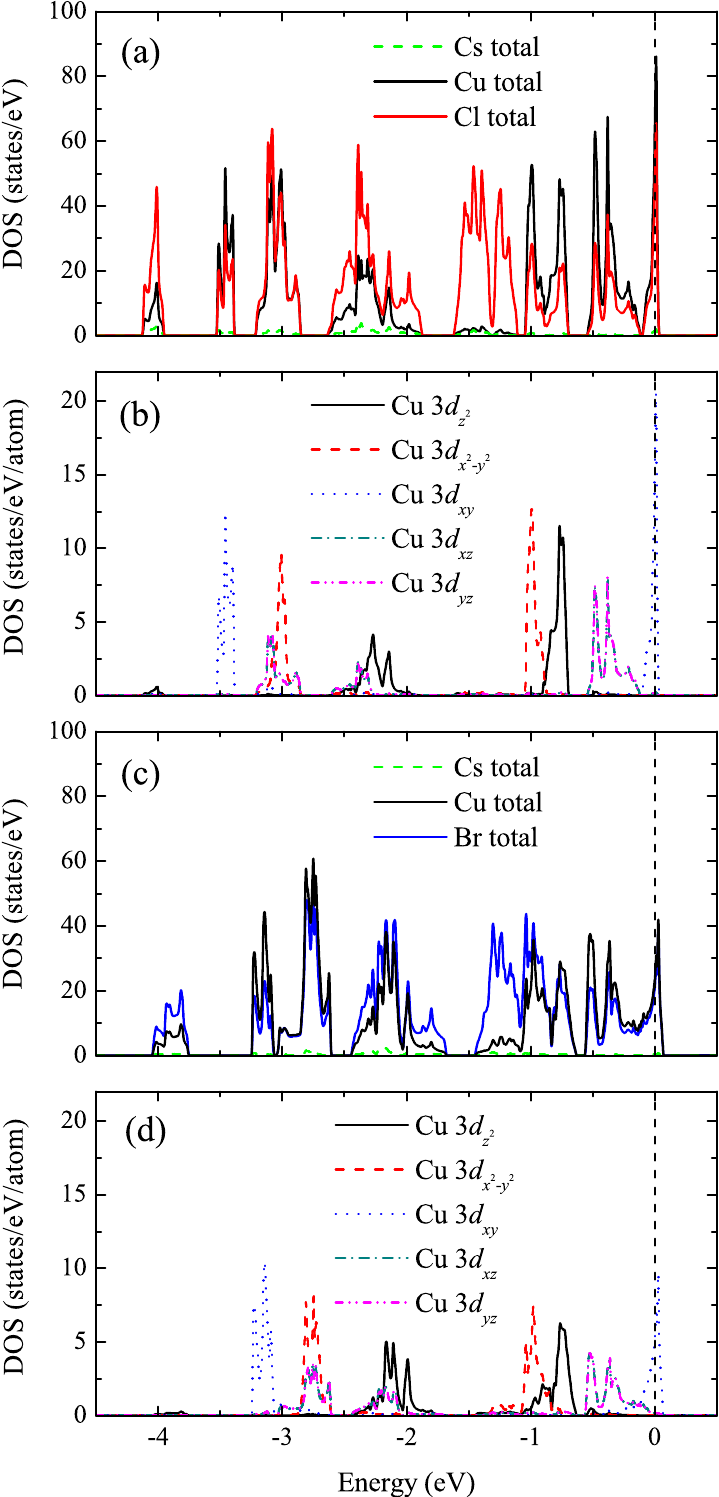}
\caption{(Color online) Atomic species resolved densities of states (DOS) calculated
  from the experimental crystal structures for (a) Cs$_2$CuCl$_4$ and
  (c) Cs$_2$CuBr$_4$ and the orbital projected densities of states of
  Cu $3d$ for (b) Cs$_2$CuCl$_4$ and (d) Cs$_2$CuBr$_4$. Energy is
  measured relative to the Fermi level $E_\text{F}$. }
\label{proj_DOS}
\end{center}
\end{figure}

\begin{figure}
\begin{center}
\subfigure {\includegraphics[width=0.45\textwidth]{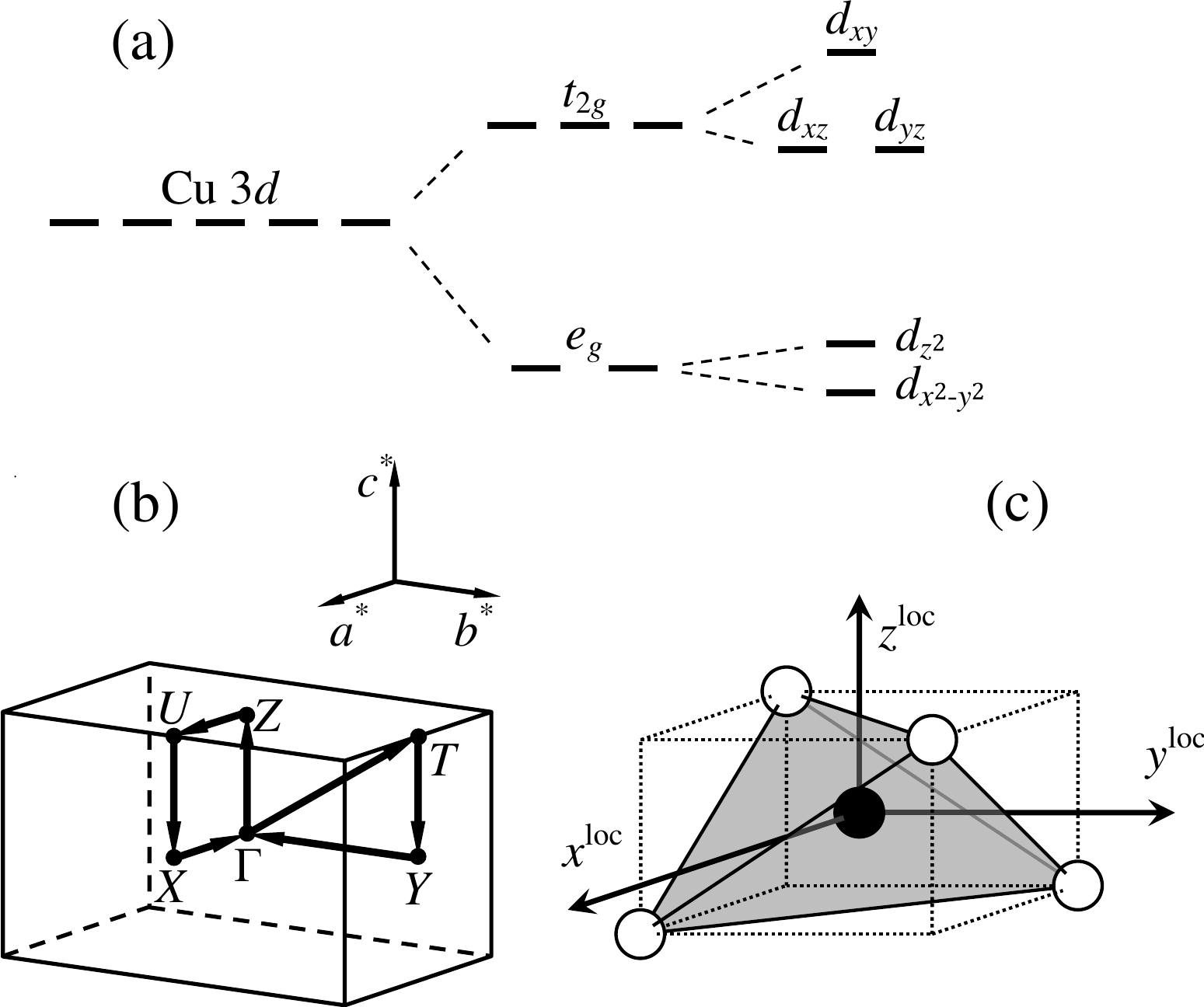}}
\caption{(a) A schematic splitting of Cu $3d$ orbitals in the crystal
  field.  (b) The path in the Cs$_2$CuCl$_4$/Cs$_2$CuBr$_4$ Brillouin
  zone for bandstructure calculations (Figs.~\ref{El_str_cl} and
  \ref{El_str_br}).  (c) The local reference frame of a Cu atom, in
  which its orbital projected density of states is defined.
}\label{misc}
\end{center}
\end{figure}

Fig.~\ref{proj_DOS} displays the density of states (DOS) for
Cs$_2$CuCl$_4$ and Cs$_2$CuBr$_4$ obtained using the experimental
structure.  In both compounds, the hybridized Cu $3d$ and Cl $3p$/Br
$4p$ bands occupy the energy range between about -4~eV and 0~eV [see,
  for instance, the atomic DOS in Figs.~\ref{proj_DOS}~(a) and
  (c)]. At the Fermi level, the Cu $3d_{xy}$ states are half-filled as
Cu is in a $3d^9$ configuration.
There are four Cu $3d_{xy}$ bands in the bandstructure, which
corresponds to the number of Cu atoms per unit cell.  A gap of
approximately 4~eV separates the Cu and $X$ ($X=\text{Cl}$, Br) bands
from the next unoccupied states [not shown in Figs.~\ref{proj_DOS}~(a)
  and (c)], which have significant Cs contribution.

Almost no contribution from Cs atoms to the DOS near the Fermi level
is observed indicating a negligible hybridization of Cu with Cs.  In
particular, this indicates that exchange coupling $J$ along the Cu
chains in the $b$ direction arises from the Cu-$X$-$X$-Cu
hybridization.

The Cu and $X$ band manifold is an assembly of bonding states in the
interval between -4~eV and -2~eV and antibonding states in the
interval from -2~eV up to the Fermi level. The Cu antibonding states
are split by the crystal field generated by $X^{-}$ ions surrounding a
Cu$^{2+}$ ion into the energetically lower Cu $e_g$ doublet
($d_{x^2-y^2}$ and $d_{z^2}$) and the energetically higher Cu $t_{2g}$
triplet ($d_{xy}$, $d_{xz}$ and $d_{yz}$).  Due to the
Jahn-Teller-like uniaxial distortion of the tetrahedron, the $t_{2g}$
triplet is further split into the degenerate $d_{xz}$/$d_{yz}$ states
and the half-filled $d_{xy}$ states.  This splitting is schematically
illustrated in Fig.~\ref{misc}~(a), and the orbital projected
densities of Cu $3d$ states for the experimental Cs$_2$CuCl$_4$ and
Cs$_2$CuBr$_4$ structures are presented in Figs.~\ref{proj_DOS}~(b)
and (d), respectively.  Note that the orbital designation is given
according to the local reference frame of the Cu$X_4$ tetrahedron as
shown in Fig.~\ref{misc}~(c).

\begin{figure}
\begin{center}
\subfigure {\includegraphics[width=0.48\textwidth]{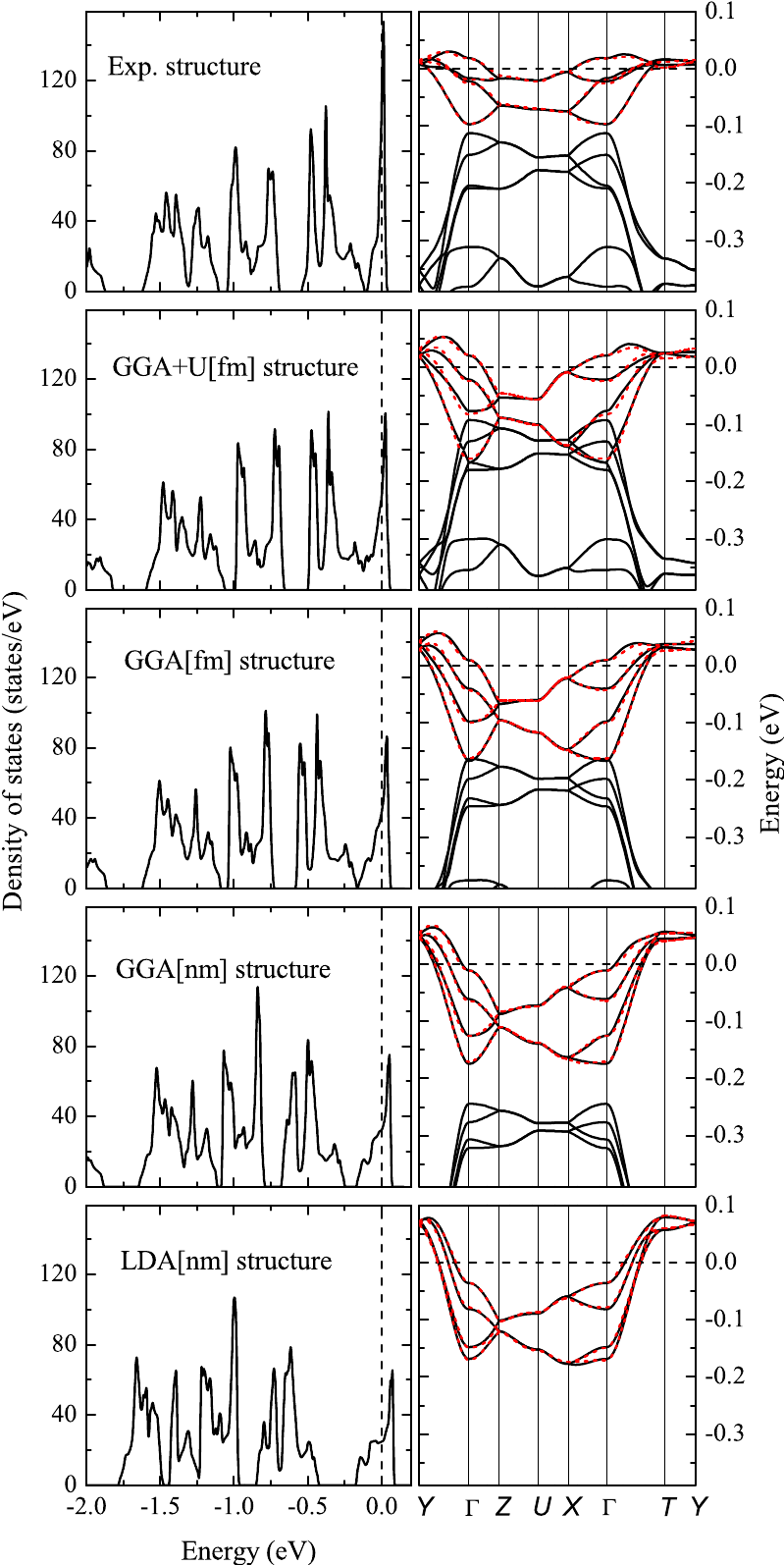}}
\caption{(Color online) DOS and bandstructures for various Cs$_2$CuCl$_4$ structures
  (specified by the panel captions). In the bandstructure plots, the
  DFT calculated bands are in black (solid) lines and the tight-binding
  fits are in red (dashed) lines. The bandstructure path in the
  Brillouin zone is shown in Fig.~\ref{misc}~(b). Energy is again
  measured relative to the Fermi level $E_\text{F}$.}\label{El_str_cl}
\end{center}
\end{figure}

\begin{figure}
\begin{center}
\subfigure {\includegraphics[width=0.48\textwidth]{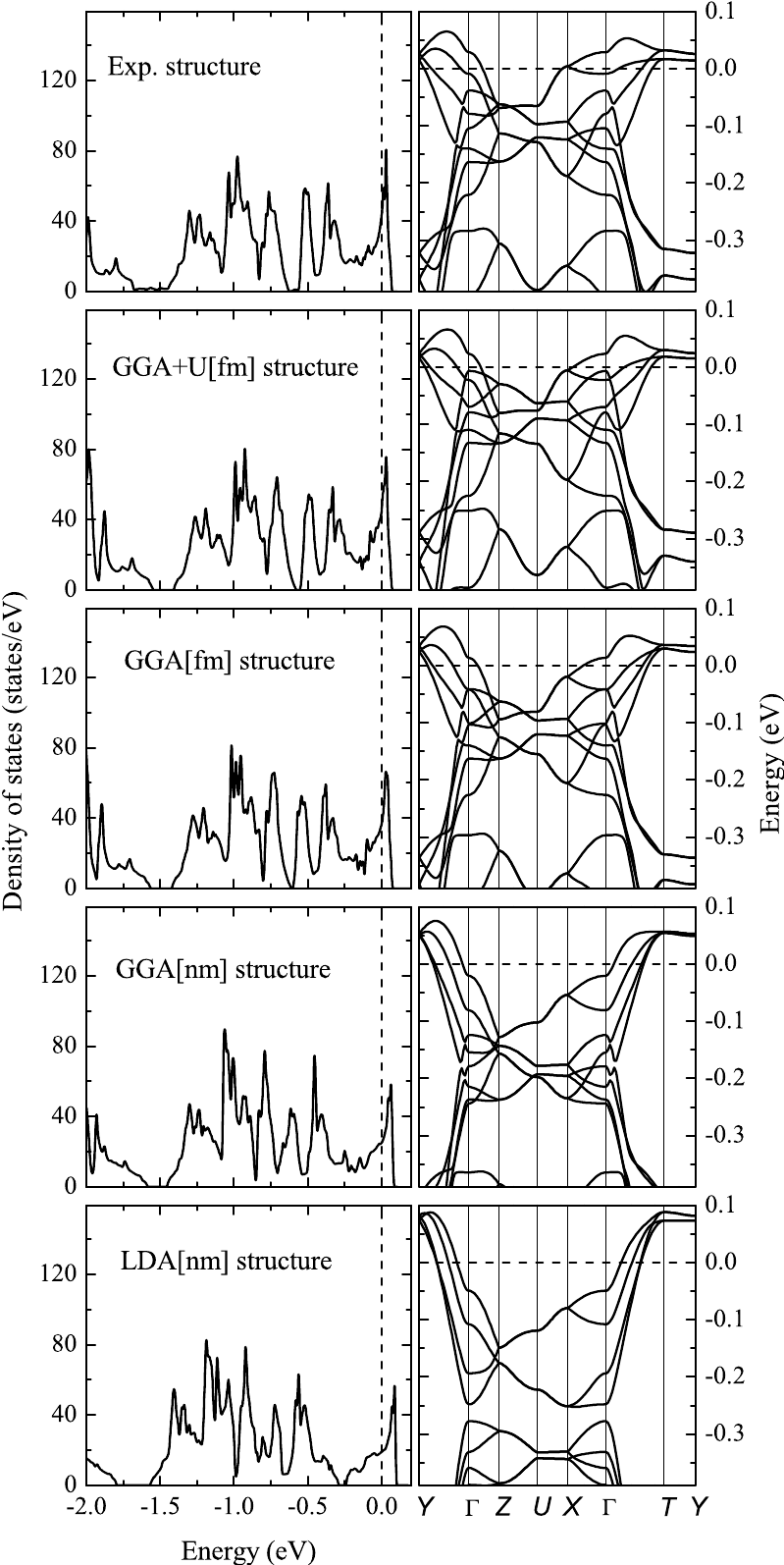}}
\caption{DOS and bandstructures of various Cs$_2$CuBr$_4$
  structures (specified by the panel captions). The bandstructure path in the Brillouin zone is shown in
  Fig.~\ref{misc}~(b).}\label{El_str_br}
\end{center}
\end{figure}

In Figs.~\ref{El_str_cl} and \ref{El_str_br}, we present the total DOS
and bandstructures for the experimental as well as relaxed crystal
structures ({\fmggau}, {\fmgga}, {\nmgga}, {\nmlda}) of Cs$_2$CuCl$_4$ and
Cs$_2$CuBr$_4$, respectively.
In contrast to Cs$_2$CuCl$_4$, the $t_{2g}$ states in Cs$_2$CuBr$_4$
are strongly hybridizing, which is indicated by the non-separable
character of the overlap of the Cs$_2$CuBr$_4$ $d_{xy}$ and
$d_{xz}$/$d_{yz}$ bands in the bandstructure.

We now focus on the comparison between the electronic structures of
Cs$_2$CuCl$_4$ and Cs$_2$CuBr$_4$ calculated from different crystal
structures.
In both compounds, the choice of the relaxation scheme determines the
degree of separation between the $d_{xy}$ and $d_{xz}$/$d_{yz}$ bands
and the dispersion of the $d_{xy}$ band.  The separation between the
$d_{xy}$ and $d_{xz}$/$d_{yz}$ bands is reduced in the {\nmgga} crystal
structure, compared to the {\nmlda} one, and keeps reducing within the
GGA series as the value of the Cu magnetic moment, associated with a
given relaxation functional, gets larger ({\fmgga}, {\fmggau}).  This
trend is better seen in Cs$_2$CuCl$_4$ where the $d_{xy}$ band is
separated by a gap (except for the {\fmggau} structure). We relate the
variation of the degree of separation to the variation of the
Jahn-Teller-like distortion of the Cu$X_4$ tetrahedron. Our assumption
is supported by the structural analysis in Tables~\ref{tetra_Cl}
and~\ref{tetra_Br} as one observes there that the structures
possessing more distorted tetrahedra demonstrate larger $d_{xy}$ and
$d_{xz}$/$d_{yz}$ band separation in the electronic structure.

Analysis of the $d_{xy}$ band dispersion shows that the $d_{xy}$
bandwidth does not change significantly among the Cs$_2$CuBr$_4$
structures including the experimental one (Fig.~\ref{El_str_br}).  In
Cs$_2$CuCl$_4$, the $d_{xy}$ bandwidth of the experimental crystal
structure is significantly narrower compared to that of the relaxed
crystal structures, where it stays nearly uniform
(Fig.~\ref{El_str_cl}).  The shape of the $d_{xy}$ bands varies in
details from structure to structure in both compounds. This shape
variation is accompanied by a variation of the density of states
distribution within the $d_{xy}$ orbital, which is easier to quantify.
Thus, in both compounds, the weight of the $d_{xy}$ DOS gets shifted
closer to the Fermi energy in the {\nmlda}, {\nmgga}, {\fmgga}, {\fmggau}
crystal structures sequence, which results in a more peaked appearance
of the $d_{xy}$ DOS.  The DOS at the Fermi level increases
correspondingly. In the experimental Cs$_2$CuCl$_4$ structure, the DOS
at the Fermi level is extremely sharp also due to the flatness of the
$d_{xy}$ band.

\subsection{ (e) TB model for  Cs$_2$CuCl$_4$}
The changes observed among the various electronic structures of
Cs$_2$CuCl$_4$ and Cs$_2$CuBr$_4$ can also be discussed on a
quantitative level by mapping bandstructures to a tight-binding (TB)
Hamiltonian:
\begin{equation}
 H^{\rm TB} = \mu\sum_i \hat{c}^{\dagger}_{i,\sigma}\hat{c}_{i,\sigma}
- \sum_{\scriptsize
\begin{array}{c}
<i,j>,\sigma\\
i\neq j
\end{array}
} \left( t_{ij}
\hat{c}^{\dagger}_{i,\sigma}\hat{c}_{j,\sigma} + {\rm h.\, c.}
\right),\label{H_TB}
\end{equation}
where $\hat{c}^{\dagger}_{i,\sigma}$ and $\hat{c}_{i,\sigma}$ are,
respectively, the creation and annihilation operators of electrons on
site $i$ and with spin $\sigma$, $t_{ij}$ denotes the hopping
integrals and h. c. denotes Hermitian conjugate terms.

Since in Cs$_2$CuCl$_4$ the Cu $3d_{xy}$ bands at the Fermi level are
well separated from the rest of the $t_{2g}$ manifold, its low-energy
properties can be described by a one-band TB model.
The hopping integrals are evaluated by Fourier-transforming the
Hamiltonian Eq.~(\ref{H_TB}) and mapping its eigenvalues to the four
DFT Cu $3d_{xy}$ bands.

In Cs$_2$CuBr$_4$, the TB procedure is complicated by the
hybridization of the overlapping $d_{xy}$ and $d_{xz}/d_{yz}$
bands. In this case, a three-band TB model is required. Since the
three-band TB model includes new electronic degrees of freedom due to
electron hopping between different types of orbitals, the number of
independent hopping integrals in this model is considerably larger,
compared to the one-band case. With an increasing number of model
parameters, the TB parametrization by means of fitting becomes less
reliable, and therefore we do not apply this method to Cs$_2$CuBr$_4$
in the present work. The rest of this section will be dedicated to the
discussion of the Cs$_2$CuCl$_4$ single-band TB models.

\begin{table}
\caption{The TB model parameters in meV for the Cs$_2$CuCl$_4$
  Cu~3$d_{xy}$ band, calculated from the various Cs$_2$CuCl$_4$
  crystal structures. The hopping integral index corresponds to the
  order of the neighbor. The less important interaction pathways, not
  present in Fig.~\ref{crystal}~(b), are shown in
  Fig.~\ref{pathways}~(a) (Appendix B).} \label{TB_models}
\begin{tabular}{lrrrrr}
\hline \hline
& {\nmlda} & {\nmgga} & {\fmgga} & {\fmggau} & exp. \\
\hline
$t$      &-44.9\;\;\;\;&-35.9\;\;&-27.4\;\;\;\;&-21.7\;\;\;\;\;&-11.0\;\\
$t'$     & 12.5\;\;\;\;&-13.6\;\;& 14.0\;\;\;\;& 14.4\;\;\;\;\;&  6.7\;\\
$t''$    & -1.4\;\;\;\;& -4.5\;\;& -6.0\;\;\;\;& -6.8\;\;\;\;\;& -6.3\;\\
$t_1$    &  6.3\;\;\;\;& -7.4\;\;& -6.3\;\;\;\;& -3.6\;\;\;\;\;& -3.9\;\\
$t_3$    & -9.5\;\;\;\;&  8.4\;\;&  7.5\;\;\;\;&  7.5\;\;\;\;\;&  8.2\;\\
$t_7$    &  2.4\;\;\;\;&  2.3\;\;&  2.6\;\;\;\;&  2.9\;\;\;\;\;&  2.3\;\\
$t_8$    & -2.2\;\;\;\;& -2.5\;\;& -2.7\;\;\;\;& -3.0\;\;\;\;\;&  3.6\;\\
$t_6$    & -2.4\;\;\;\;& -2.8\;\;& -3.0\;\;\;\;& -2.7\;\;\;\;\;&  1.7\;\\
$t_{14}$ &  0.6\;\;\;\;&  0.9\;\;&  1.1\;\;\;\;&  1.4\;\;\;\;\;&  1.6\;\\
$t_{18}$ &  0.0\;\;\;\;&  0.2\;\;&  0.3\;\;\;\;&  0.6\;\;\;\;\;& -0.1\;\\
$t_{22}$ & -4.8\;\;\;\;& -5.1\;\;& -4.5\;\;\;\;& -4.4\;\;\;\;\;& -2.4\;\\
$\mu$    &-15.3\;\;\;\;&-17.3\;\;&-17.6\;\;\;\;&-17.8\;\;\;\;\;&-11.7\;\\
\hline \hline
\end{tabular}
\end{table}

Though it was experimentally established \cite{Coldea02} that the spin
interactions in Cs$_2$CuCl$_4$ are predominantly within the Cu layer
along the $J$ and $J'$ paths [see Fig.~\ref{crystal}~(b)], we observe
that the electronic behavior modeled by the TB Hamiltonian show
non-negligible interlayer hopping terms.  We find that the $d_{xy}$
bands of all the Cs$_2$CuCl$_4$ structures can only be satisfactorily
described with a minimal model that includes five hopping integrals.
Three of them are the intralayer hoppings $t$ and $t'$ and the
interlayer hopping $t''$ ($t$, $t'$, $t''$ correspond to the
interaction paths $J$, $J'$, $J''$, {\it etc.}), which have been
considered in previous studies in the framework of spin
Hamiltonians. The two new hopping parameters are the interlayer $t_1$
and $t_3$, also shown in Fig.~\ref{crystal}~(b), their indices
indicating the order of the interacting Cu neighbor.
Note that $t$, $t'$ and $t''$ as well as $J$, $J'$ and $J''$ correspond to fifth, fourth and second nearest neighbour
in the structure, respectively, while for all other $t_i$ and $J_i$ the index corresponds
to the order of the neighbour. 
 In
Table~\ref{TB_models}, we present the values of relevant hopping
integrals for the various Cs$_2$CuCl$_4$ structures.  We observe that
$t_1$ and $t_3$ are in many cases comparable in magnitude to the
hopping integral $t'$.  We also note that the presence of
non-negligible interlayer couplings is revealed by a significant
dispersion of the $d_{xy}$ bands along the $a^*$ direction in
$k$-space (the $X-\Gamma$ path in Fig.~\ref{El_str_cl}). Additionally,
we confirmed the importance of $t_1$ and $t_3$ by performing NMTO
downfolding calculations \cite{Andersen00} for some of the relaxed
crystal structures.

The total number of TB hopping integrals considered for an accurate
description of the DFT bandstructure of Cs$_2$CuCl$_4$ amounts to 11,
plus the onsite energy $\mu$ (Table~\ref{TB_models}).
Fig.~\ref{El_str_cl} displays in dashed red lines the TB fit to the
DFT Cu 3$d_{xy}$ bands, which are in solid black lines. Interaction
pathways for the first seven hopping integrals are shown schematically
in Figs.~\ref{crystal}~(b) and \ref{pathways}~(a) (in Appendix A).

Comparing the TB models for different Cs$_2$CuCl$_4$ structures, we
observe that the dominant hopping integral $t$ has the smallest
absolute value for the experimental crystal structure and increases in
the GGA crystal structure series with a decrease of the Cu magnetic
moment during relaxation ({\fmggau}, {\fmgga}, {\nmgga}).  It has a maximum
value in the {\nmlda} structure. At the same time, the second important
coupling $t'$ does not change significantly among the relaxed
structures, but is by a factor of two smaller in the experimental
crystal structure.  The interlayer coupling $t_3$ is stable for all
the structures, while $t''$ and $t_1$ vary considerably.

We associate the variation of the hopping integral $t$ in different
Cs$_2$CuCl$_4$ crystal structures with the variation of the CuCl$_4$
tetrahedron geometry and, in particular, with the variation of the
angle $\alpha$ in the Cu-Cl(3)-Cl(3)-Cu interaction path. In the
relaxed Cs$_2$CuCl$_4$ crystal structures, larger $t$ values
correspond to larger values of $\alpha$ and larger tetrahedron
distortions.  However, the experimental structure does not follow this
rule; while the angles of the experimental and {\fmggau} relaxed
structures are very similar, the $t$ values differ by a factor of two.
The reason why the experimental structure deviates from the relaxed
structures might be the influence of other tetrahedron parameters,
such as details of the Cu-Cl bond distances.

The observed relation between the dominant hopping integral $t$ and
the angle $\alpha$ is reasonable since $\alpha$ is the defining angle
for the Cu $3d_{xy}$-Cl $3p$-Cl $3p$-Cu $3d_{xy}$ hybridization.  By
considering perturbation theory on the onsite Coulomb repulsion $U$ up
to the second order, the effective Cu-Cu superexchange coupling can be
obtained from $t$ as $J=\frac{4t^2}{U}$.  Then, the relation between
$t$ and $\alpha$ fulfills the Kanamori-Goodenough
rule~\cite{Goodenough58, Kanamori59}, stating that $J$ reaches a
maximum when the cation-anion-cation angle equals 180$^\circ$. In the
present case of the cation-anion-anion-cation (Cu-Cl-Cl-Cu) bridge,
the four atoms get aligned along a straight line when $\alpha$
increases.

\section{III. Exchange Integrals}
\subsection{ (a) Computational details}
The exchange coupling integrals $J_{ij}$ of the Heisenberg Hamiltonian,
\begin{equation}
H_{\rm H} = \sum_{<i,j>}J_{ij}\boldsymbol{S}_i\boldsymbol{S}_j, \label{Heisenberg}
\end{equation}
can be obtained by means of DFT spin-resolved total energy
calculations~\cite{Kunes05}. Considering the differences between the energies of the
ferromagnetic configuration and various antiferromagnetic spin
configurations, $E^{\rm FM}-E^\text{AFM}_i$, one derives a set of
coupled equations for the couplings $J_{ij}$.  Following
Eq.~(\ref{Heisenberg}), antiferromagnetic exchange corresponds to
positive $J_{ij}$.

\begin{figure*}
\begin{center}
\subfigure {\includegraphics[width=0.95\textwidth]{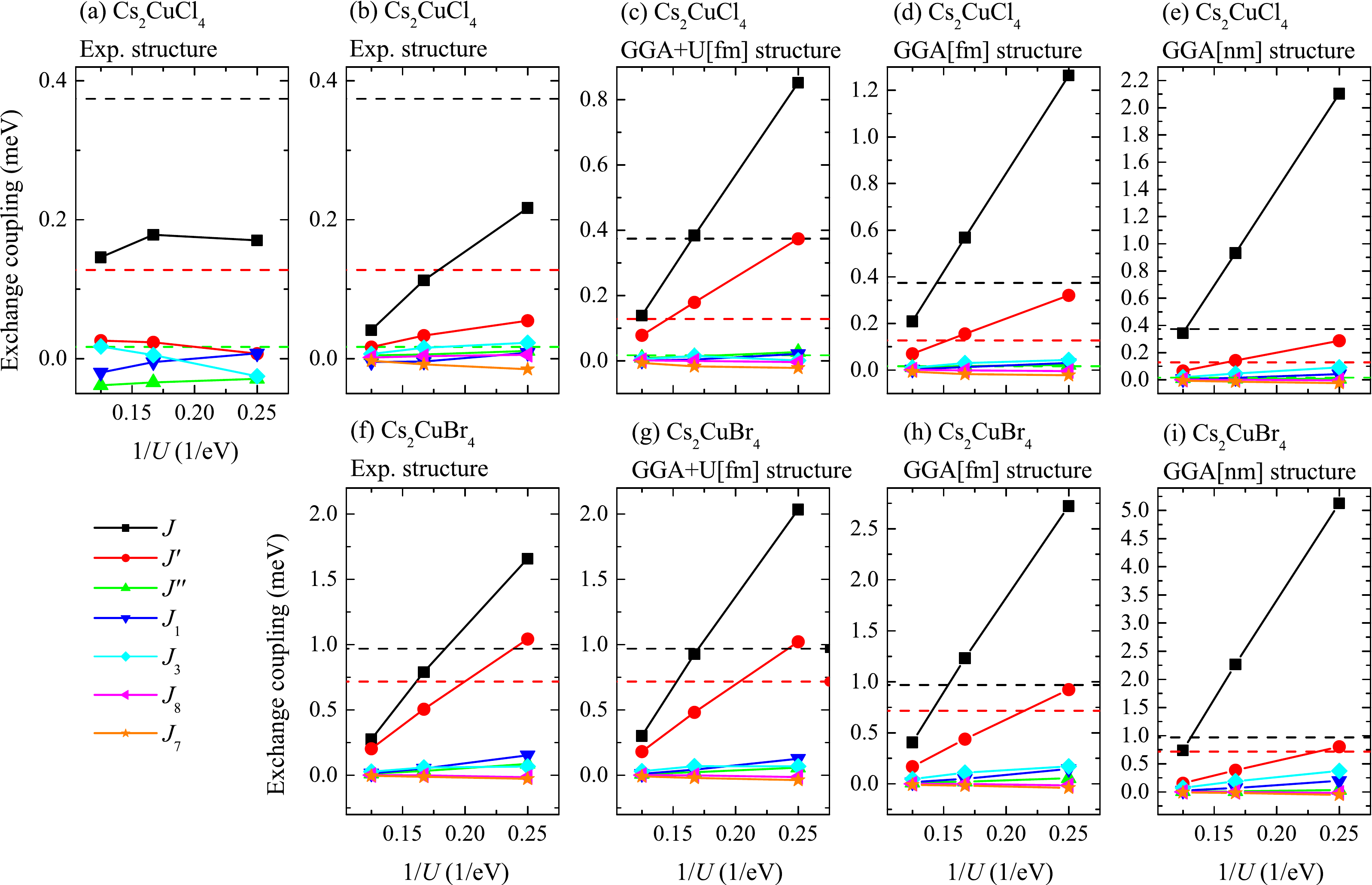}}
\caption{(Color online) The spin exchange coupling constants as functions of $1/U$
  for (a) the Cs$_2$CuCl$_4$ experimental structure calculated with
  the atomic limit version of the GGA+U, (b) the Cs$_2$CuCl$_4$
  experimental structure (c) the Cs$_2$CuCl$_4$ {\fmggau} relaxed
  structure, (d) the Cs$_2$CuCl$_4$ {\fmgga} relaxed structure, (e) the
  Cs$_2$CuCl$_4$ {\nmgga} relaxed structure, (f) the Cs$_2$CuBr$_4$
  experimental structure, (g) the Cs$_2$CuBr$_4$ {\fmggau} relaxed
  structure, (h) the Cs$_2$CuBr$_4$ {\fmgga} relaxed structure, (i) the
  Cs$_2$CuBr$_4$ {\nmgga} relaxed structure.  When not specified
  otherwise, the exchange couplings are obtained with the around mean
  field version of the GGA+U.  The three sets of exchange couplings
  correspond to $U=8$, $6$ and $4$~eV.  Dashed lines mark the
  experimentally determined values of $J$, $J'$ and $J''$.  }\label{J}
\end{center}
\end{figure*}

For Cs$_2$CuCl$_4$ and Cs$_2$CuBr$_4$, we take into account seven
important couplings corresponding to the first seven hopping integrals
in Table~\ref{TB_models}, which are $J$, $J'$,
$J''_\text{eff}=J''+J_6$, $J_1$, $J_3$, $J_7$ and $J_8$ [see
  Figs.~\ref{crystal}~(b) and \ref{pathways}~(a)]. This choice defines
the number of antiferromagnetic configurations to be calculated. The
combined coupling $J''_\text{eff}$ is introduced because the
considered unit cell (which is a $2\times2\times1$ supercell) does not
allow a separate calculation of the couplings $J''$ and $J_6$ but only
their calculation as a sum. Since $J''$ is presumably larger than
$J_6$, $J''_\text{eff}$ gives an approximate value of $J''$.

The choice of the supercell is dictated by the peculiarities of the
Cs$_2$CuCl$_4$ and Cs$_2$CuBr$_4$ magnetic sublattices. First, since
the candidate for the largest coupling $J$ connects Cu atoms that
belong to adjacent primitive unit cells in Cs$_2$CuCl$_4$ and
Cs$_2$CuBr$_4$, the primitive unit cell has to be doubled in the $b$
direction. Otherwise, $J$ would be always canceled in any
$E^\text{FM}-E^\text{AFM}_i$ difference. Also, in order to be able to
discern the inequivalent couplings $J_1$ and $J_3$, we double the unit
cell once more in the $a$ direction and thus end up with a
$2\times2\times1$ supercell. In the supercell, we set eight out of 16
Cu atoms inequivalent in order to be able to arrange the required
seven antiferromagnetic configurations within the same unit cell space
group, which is {\it P}-1. It is important to stay within the same
space group during total energy calculations for Cs$_2$CuCl$_4$ and
Cs$_2$CuBr$_4$ as in these compounds the exchange couplings are small
and integration over differently sampled Brillouin zones can affect
the accuracy of the results.

The seven antiferromagnetic spin configurations, together with the set
of coupled equations for the exchange couplings, are presented in
Appendix C. The total energy calculations were performed with the FPLO
code. Test calculations with Wien2k confirm the results. In the FPLO
code, we chose a $5\times4\times3$ mesh of $k$-points for the
supercell Brillouin zone integration and kept other settings at
default. The scheme to compute magnetic exchange for a given structure
consisted of a series of total energy calculations within the AMF
version of the GGA+U, with $U = 4$, 6 and 8~eV and $J_\text{H}=1$~eV
in all cases. This scheme was applied to the experimental structures
of Cs$_2$CuCl$_4$ and Cs$_2$CuBr$_4$ and also to the {\nmgga}, {\fmgga}
and {\fmggau} relaxed structures of the two compounds.

\subsection{ (b) Results}

The calculated exchange couplings are plotted in Fig.~\ref{J}, as a
function of $1/U$. For an easy comparison, we also mark by horizontal
dashed lines of corresponding colors the experimentally determined
values of $J$, $J'$ and $J''$ for Cs$_2$CuCl$_4$ \cite{Coldea02} and
of $J$ and $J'$ for Cs$_2$CuBr$_4$ \cite{Ono05_2}. Additionally, the
exchange coupling values are provided in Appendix D.

We first compare the Heisenberg models of Cs$_2$CuCl$_4$, derived for
the experimental as well as relaxed crystal structures
[Figs.~\ref{J}~(a)-(e)].  The common feature for all these structures
is that their spin models are two-dimensional, with $J$ being the
leading interaction followed by $J'$.

For the experimental structure [Fig.~\ref{J}~(b)], the ratio $J'/J$ is
similar to the experimentally determined
ratio~\cite{Coldea02,Foyevtsova09}, while the absolute values of $J$
and $J'$, which are calculated in the present work,  are off by a factor of almost 3.  In this structure, the
two-dimensionality is less pronounced due to the interlayer
antiferromagnetic interaction $J_3$ and the ferromagnetic interaction
$J_7$, which are competing with $J'$.  The {\nmgga} structure
[Fig.~\ref{J}~(e)], on the other hand, tends to behave as a 1D rather
than 2D system as here the $J'/J$ ratio is 0.15 at $U=6$~eV.  We
assume that $U=6$~eV should be close to the $U$ value for the Cu 3$d$
electrons in the LAPW GGA+U scheme (see, for instance,
Ref.~\onlinecite{Haas07} where the authors successfully reproduce the
electric field gradients at Cu$^{2+}$ ions in a number of Cu$^{2+}$
oxides and halides by performing LDA+U calculations with $U^{\rm eff}
= U-J = 5$~eV).  The $J'/J$ ratios at $U=6$~eV in the experimental,
{\fmggau} and {\fmgga} structures are, respectively, 0.30, 0.47 and
0.27. These ratios vary slightly for other $U$ values.

The differences in the $J'/J$ ratios among the relaxed structures are
mainly due to a strong variation of the coupling $J$, which adopts the
values 0.384, 0.568, 0.932 meV for $U=6$~eV in the {\fmggau}, {\fmgga}
and {\nmgga} structures, respectively, whereas $J'$ decreases only
slightly for the same sequence of structures.

Judging by the proximity of the theoretically derived Cs$_2$CuCl$_4$
Heisenberg model (this work) to the experimentally determined one
\cite{Coldea02}, the calculations with the {\fmgga} and {\fmggau}
optimized structures provide the most satisfactory results.  In both
cases, the absolute values of $J$ and $J'$ as well as their ratios are
close to experiment for typical values of $U$ between 6 and 8~eV (for
{\fmggau} $U\approx 6$~eV yields the best agreement, while in the case
of {\fmgga} interpolation of the results between 6 and 8~eV results in
almost perfect agreement with experiment for $U\approx
7$~eV). Moreover we find non-negligible interlayer couplings as shown
in Appendix D (bold-face values).

For the spin models calculated with the relaxed Cs$_2$CuCl$_4$
structures, the ratios of antiferromagnetic exchange couplings are in
good agreement with the ratios of the corresponding squared hopping
integrals.  This is an indication that ferromagnetic contributions to
these couplings are small.

In contrast to Cs$_2$CuCl$_4$, the Cs$_2$CuBr$_4$ effective spin model
derived using the experimental crystal structure [Fig.~\ref{J}~(f)]
agrees well with the experimentally estimated model parameters in the
interval of $U$ values around 6~eV.  We obtain that Cs$_2$CuBr$_4$ is
a two-dimensional system, with dominant antiferromagnetic couplings
$J$ and $J'$ and considerably smaller interlayer couplings. At
$U=6$~eV, the obtained $J'/J$ ratio equals 0.64, which compares well
with the experimental result of 0.74.  Also, the model based on the
{\fmggau} relaxed structure [Fig.~\ref{J}~(g)] gives similar results
for the exchange parameters.  The similarity between these two models
is another feature that distinguishes Cs$_2$CuBr$_4$ from
Cs$_2$CuCl$_4$. In the latter case, as seen above, the exchange
couplings of the experimental model [Fig.~\ref{J}~(b)] are
considerably smaller than those of the {\fmggau} relaxation model
[Fig.~\ref{J}~(c)].

Overall, the Heisenberg models for Cs$_2$CuBr$_4$ obtained within the
GGA relaxation series [Figs.~\ref{J}~(g)-(i)] follow the same behavior
as the corresponding Cs$_2$CuCl$_4$ models, which is characterized by
increasing $J$ and decreasing $J'/J$ in structures that have been
relaxed with a smaller Cu magnetic moment.

To conclude this section, we briefly comment on the performance of the
atomic limit (AL) version \cite{Anisimov93} of the GGA+U
exchange-correlation functional, which is an alternative to the around
mean field version. The exchange couplings of the experimental
Cs$_2$CuCl$_4$ structure calculated with the AL double counting
correction differ considerably from those obtained with the AMF double
counting correction [compare panels (b) and (c) of Fig.~\ref{J}]. The
AL calculated exchange couplings $J$ and $J'$ do not behave linearly
with $1/U$, as expected from the $J=4t^2/U$ relation, valid for these
antiferromagnetic couplings. Therefore, we restrict ourselves in the
remaining discussion to calculations with the AMF double counting
correction.

\section{IV. Discussion}
From our analysis of the structural and electronic properties of
Cs$_2$CuCl$_4$ and Cs$_2$CuBr$_4$ we obtained the following results:
Within the sequence {\nmlda}, {\nmgga}, {\fmgga} and {\fmggau} of functionals
used for the structural optimization, the Jahn-Teller-like distortion of
the Cu$X_4$ tetrahedron is reduced, which is accompanied by an increase
of the Cu-$X(3)$-$X(3)$ angle $\alpha$ in the $J$ superexchange path.
These structural changes lead to a considerable variation of the exchange
coupling $J$ and the corresponding hopping integral $t$.

\subsection{(a) Importance of magnetism and electronic correlations}
We find that the choice of the functional used in the structure
optimization is crucial for the correct modeling of the properties of
these two compounds, especially in the case of Cs$_2$CuCl$_4$.  First
of all, opening of a gap at the Fermi level by introducing various
magnetic structures seems to be a necessary ingredient in the
structure optimization, as can be seen from the relatively bad
performance of the {\nmlda} and {\nmgga} optimizations. Since in these
schemes the system is gapless, which is in stark contrast to the true
ground state of Cs$_2$CuCl$_4$ observed experimentally, the strong
instability at the Fermi level (visible in the peak in the DOS) has to
be partially relieved by forcing an unphysically strong structural
distortion, leading to an improper determination of lattice structure.
Furthermore, a change of the size of the gap through onsite
correlations and corresponding localization of the Cu magnetic moments
also influences the results of the structure optimization.  Although
one might have expected a better performance of the spin-dependent
optimization schemes (due to the magnetic nature of the compounds),
the strong impact on the calculated exchange couplings is rather
unexpected.  The reason is presumably the small values of the exchange
constants.

\subsection{(b) Comments on the quality of the experimentally determined Cs$_2$CuCl$_4$ and Cs$_2$CuBr$_4$
crystal structures} Comparison of electronic behavior and effective
models of the various relaxed Cs$_2$CuCl$_4$ structures with those of
the experimental structure indicates that the experimental
determination of the Cs$_2$CuCl$_4$ crystal structure
\cite{Bailleul91} was probably not sufficiently accurate, at least, as
far as calculations of microscopic models are concerned.  This would
explain the huge differences between our derived Cs$_2$CuCl$_4$
Heisenberg model when the experimental crystal structure is
considered, and the model Coldea {\it et al.}~\cite{Coldea02} obtained
from fitting to neutron data and which was corroborated by a number of
studies~\cite{Zheng05,Veillette05,Starykh07,Starykh10}.

Accurate determination of the Cs$_2$CuCl$_4$ crystal structure might
be complicated due to the presence of non-stoichiometric hydrogen
containing compounds, presumably HCl or HO$_2$, detected in this
material and not detected in
Cs$_2$CuBr$_4$~\cite{Bailleul91,Assmus10}.

The experimentally determined Cs$_2$CuBr$_4$ crystal structure, on the
other hand, is accurate enough.  We suggest therefore that
Cs$_2$CuBr$_4$ can be regarded as a reference system for choosing the
relaxation scheme, suitable also for describing Cs$_2$CuCl$_4$. We
conclude from the data analysis and from the physical considerations
that such a relaxation scheme is the {\fmgga}({\fmggau}), with 0 $\leq U \leq 6$~eV.

\subsection{(c) Comparison with the experimentally determined microscopic models}
In view of the arguments presented above, one should refer to the
results obtained with either {\fmgga} or {\fmggau} relaxed structures
when discussing the realistic spin model for Cs$_2$CuCl$_4$.  Our
calculations confirm for both systems that their spin models are 2D,
with the intraplane $J$ and $J'$ being the leading interactions, and
Cs$_2$CuBr$_4$ showing a higher degree of frustration than
Cs$_2$CuCl$_4$.
The model from Coldea {\it et al.}~\cite{Coldea02}, with $J=0.374$~eV
and $J'/J=0.34$, is close to our DFT models derived with the {\fmgga}
and {\fmggau} relaxed structures and remains a valid model for
Cs$_2$CuCl$_4$.  In Cs$_2$CuBr$_4$, where the experimental crystal
structure seems to be much more reliable, we can propose an
approximate DFT model, obtained as a generalization of the models of
Figs.~\ref{J}~(f) and (g) for $U\approx 6$~eV: $0.8 \lesssim J
\lesssim 0.9$, $0.5 \lesssim J'/J \lesssim 0.65$. This model is quite
close to the model by Ono {\it et al.}~\cite{Ono05_2}, with
$J=0.97$~eV and $J'/J = 0.74$.

Additionally, in view of the recent theoretical studies by Starykh
{\it et al.}~\cite{Starykh07,Starykh10}, who demonstrated the
important role of the relatively weak Dzyaloshinskii-Moriya
interaction and interlayer coupling in determining the magnetic
behavior of Cs$_2$CuCl$_4$, our results for the effective
Cs$_2$CuCl$_4$ model reveal a possible relevance of a number of
interlayer couplings, besides $J''$, which are of comparable strength
($J_1$, $J_3$ and $J_7$). Also, we find the next-nearest chain
neighbor hopping integral $t_{14}$ to be quite large, thus supporting
another suggestion by these authors.

Finally, it is important to emphasize that the electronic TB models
for the two compounds involve sizable interlayer hoppings, comparable
with the intralayer ones, so that in terms of electronic degrees of
freedom the two compounds have 3D behavior.  In the case of
Cs$_2$CuBr$_4$, this can be seen from the large dispersion of the
bands along the $a^*$ direction in $k$-space, even though we didn't
derive a TB model for this material.

\subsection{(d) Structural relaxation}
Finally, we comment on our decision to perform structural relaxations
with fixed lattice constants and using the room temperature data. It
is generally known that the GGA tends to overestimate the unit cell
volume while the LDA underestimates it. Therefore, we preferred to
rely on the experimental lattice constants, which are usually
determined with high accuracy.  We performed a test {\nmgga} relaxation
of the Cs$_2$CuCl$_4$ crystal structure with lattice constants
measured at $0.3$ K \cite{Coldea02} and found that the electronic
structure and the TB model of the resulting Cs$_2$CuCl$_4$ structure
are close to the electronic structure and the TB model of the {\nmgga}
Cs$_2$CuCl$_4$ structure, relaxed with the room temperature lattice
constants. Therefore, all calculations have been performed with the
room temperature lattice constants for both compounds.

\section{V. Summary and conclusions}

In summary, we have performed DFT calculations for Cs$_2$CuCl$_4$ and
Cs$_2$CuBr$_4$.  Our study shows that the exchange coupling constants
of these compounds exhibit a strong dependence on subtle details of
the crystal structure, especially on the geometry of the Cu$X_4$
tetrahedra. Depending on the structural model, we observe a large
variation of the derived exchange couplings and their ratios,
resulting in completely different spin model Hamiltonians.  One reason
for this unusual sensitivity are the fairly small absolute values of
the exchange couplings, with the largest coupling constant $J$ being
below 5~K.

One important motivation for our detailed study is the failure of the
experimental structure published in Ref.~\onlinecite{Bailleul91} to
correctly describe the magnetic behavior of Cs$_2$CuCl$_4$.
Calculations with the experimental structure provide too small
exchange coupling constants with fairly strong interlayer couplings in
contrast to the pronounced 2D character observed in experiment. This
indicates (together with the fairly large forces acting on the atomic
positions) that a better characterization of the Cs$_2$CuCl$_4$
crystal structure is necessary.  Only after structural optimization
with spin-dependent GGA and GGA+U functionals, we obtain an overall
good agreement with the exchange couplings obtained from
experiment~\cite{Coldea02}.

In contrast, the leading exchange couplings for Cs$_2$CuBr$_4$
obtained from our calculations are in good agreement with those
derived from experiment, independently of whether we use the
experimental structure or structures from spin-resolved
optimizations.

\section{Acknowledgments}
We would like to thank W. A\ss{}mus, F. Ritter, N. Kr\"{u}ger,
M. Lang, B. Wolf, S.S. Saxena, Y. Tanaka, and O. A. Starykh for useful
discussions. We gratefully acknowledge the Deutsche
Forschungsgemeinschaft for financial support through the SFB/TRR 49
program and the Helmholtz Association for support through HA216/EMMI.

\section{Appendix A: Structural parameters of Cs$_2$CuCl$_4$ and Cs$_2$CuBr$_4$ obtained within different optimization schemes}
Below, we provide the Cs$_2$CuCl$_4$ and Cs$_2$CuBr$_4$ relative
atomic positions obtained after structural optimization within
different schemes. For a quick reference, we also cite
Ref.~\onlinecite{Bailleul91} for the experimentally found structure of
Cs$_2$CuCl$_4$ and Ref.~\onlinecite{Morosin60} for that of
Cs$_2$CuBr$_4$.

In the case of Cs$_2$CuCl$_4$, for the experimental structure and
structures relaxed with non-spin-resolved and ferromagnetic
calculations, the lattice constants are $a=9.769$~\AA, $b=7.607$~\AA,
$c=12.381$~\AA\; and the space group is {\it Pnma}. The structures
relaxed with antiferromagnetic calculations ({\afmgga} and {\afmggau}),
for which the relaxation was constrained by the symmetry of the
$P21/c$ space group in a supercell, were found to eventually belong to
the same space group $P21/c$ but in a reduced cell, with the same unit
cell parameters as those of the original full-symmetry unit cell of
the compound.

\subsection{{\nmlda}}
\begin{tabular}{llll}
$\hspace{1.0cm}$&$\hspace{0.8cm}x\hspace{0.8cm}$&$\hspace{0.8cm}y\hspace{0.8cm}$&$\hspace{0.8cm}z\hspace{0.8cm}$\\
Cs(1)&$\hspace{0.4cm}$\;0.1322&$\hspace{0.4cm}$\;0.25 &$\hspace{0.4cm}$0.1005\\
Cs(2)&$\hspace{0.4cm}$\;0.9837 &$\hspace{0.4cm}$\;0.75 &$\hspace{0.4cm}$0.3287\\
Cu&$\hspace{0.4cm}$\;0.2322&$\hspace{0.4cm}$\;0.25 &$\hspace{0.4cm}$0.4149\\
Cl(1)&$\hspace{0.4cm}$\;0.0115 &$\hspace{0.4cm}$\;0.25 &$\hspace{0.4cm}$0.3692\\
Cl(2)&$\hspace{0.4cm}$\;0.3494&$\hspace{0.4cm}$\;0.25 &$\hspace{0.4cm}$0.5697\\
Cl(3)&$\hspace{0.4cm}$\;0.2824&$\hspace{0.4cm}$\;0.9772&$\hspace{0.4cm}$0.3654
\end{tabular}

\subsection{{\nmgga}}
\begin{tabular}{llll}
$\hspace{1.0cm}$&$\hspace{0.8cm}x\hspace{0.8cm}$&$\hspace{0.8cm}y\hspace{0.8cm}$&$\hspace{0.8cm}z\hspace{0.8cm}$\\
Cs(1)&$\hspace{0.4cm}$\;0.1329 &$\hspace{0.4cm}$\;0.25 &$\hspace{0.4cm}$0.1050\\
Cs(2)&$\hspace{0.4cm}$\;0.9864 &$\hspace{0.4cm}$\;0.75 &$\hspace{0.4cm}$0.3321\\
Cu&$\hspace{0.4cm}$\;0.2320&$\hspace{0.4cm}$\;0.25 &$\hspace{0.4cm}$0.4165\\
Cl(1)&$\hspace{0.4cm}$\;0.0045 &$\hspace{0.4cm}$\;0.25 &$\hspace{0.4cm}$0.3751\\
Cl(2)&$\hspace{0.4cm}$\;0.3507&$\hspace{0.4cm}$\;0.25 &$\hspace{0.4cm}$0.5743\\
Cl(3)&$\hspace{0.4cm}$\;0.2878&$\hspace{0.4cm}$\;0.9779&$\hspace{0.4cm}$0.3615\end{tabular}

\subsection{{\fmgga}}
\begin{tabular}{llll}
$\hspace{1.0cm}$&$\hspace{0.8cm}x\hspace{0.8cm}$&$\hspace{0.8cm}y\hspace{0.8cm}$&$\hspace{0.8cm}z\hspace{0.8cm}$\\
Cs(1)&$\hspace{0.4cm}$\;0.1318 &$\hspace{0.4cm}$\;0.25 &$\hspace{0.4cm}$0.1042\\
Cs(2)&$\hspace{0.4cm}$\;0.9903 &$\hspace{0.4cm}$\;0.75 &$\hspace{0.4cm}$0.3308\\
Cu&$\hspace{0.4cm}$\;0.2311&$\hspace{0.4cm}$\;0.25 &$\hspace{0.4cm}$0.4178\\
Cl(1)&$\hspace{0.4cm}$\;0.0039 &$\hspace{0.4cm}$\;0.25 &$\hspace{0.4cm}$0.3768\\
Cl(2)&$\hspace{0.4cm}$\;0.3479&$\hspace{0.4cm}$\;0.25 &$\hspace{0.4cm}$0.5764\\
Cl(3)&$\hspace{0.4cm}$\;0.2918&$\hspace{0.4cm}$\;0.9823&$\hspace{0.4cm}$0.3588\end{tabular}

\subsection{{\fmggau}}
\begin{tabular}{llll}
$\hspace{1.0cm}$&$\hspace{0.8cm}x\hspace{0.8cm}$&$\hspace{0.8cm}y\hspace{0.8cm}$&$\hspace{0.8cm}z\hspace{0.8cm}$\\
Cs(1)&$\hspace{0.4cm}$\;0.1321 &$\hspace{0.4cm}$\;0.25 &$\hspace{0.4cm}$0.1026\\
Cs(2)&$\hspace{0.4cm}$\;0.9948 &$\hspace{0.4cm}$\;0.75 &$\hspace{0.4cm}$0.3302\\
Cu&$\hspace{0.4cm}$\;0.2320&$\hspace{0.4cm}$\;0.25 &$\hspace{0.4cm}$0.4175\\
Cl(1)&$\hspace{0.4cm}$\;0.0043 &$\hspace{0.4cm}$\;0.25 &$\hspace{0.4cm}$0.3791\\
Cl(2)&$\hspace{0.4cm}$\;0.3442&$\hspace{0.4cm}$\;0.25 &$\hspace{0.4cm}$0.5779\\
Cl(3)&$\hspace{0.4cm}$\;0.2961&$\hspace{0.4cm}$\;0.9848&$\hspace{0.4cm}$0.3556\end{tabular}

\subsection{{\afmgga}}
\begin{tabular}{llll}
$\hspace{1.0cm}$&$\hspace{0.8cm}x\hspace{0.8cm}$&$\hspace{0.8cm}y\hspace{0.8cm}$&$\hspace{0.8cm}z\hspace{0.8cm}$\\
Cs(1)&$\hspace{0.4cm}$\;0.1317 &$\hspace{0.4cm}$\;0.2503 &$\hspace{0.4cm}$0.1044\\
Cs(2)&$\hspace{0.4cm}$\;0.9899 &$\hspace{0.4cm}$\;0.7504 &$\hspace{0.4cm}$0.3309\\
Cu&$\hspace{0.4cm}$\;0.2312&$\hspace{0.4cm}$\;0.2502 &$\hspace{0.4cm}$0.4177\\
Cl(1)&$\hspace{0.4cm}$\;0.0037 &$\hspace{0.4cm}$\;0.2497 &$\hspace{0.4cm}$0.3765\\
Cl(2)&$\hspace{0.4cm}$\;0.3483&$\hspace{0.4cm}$\;0.2507 &$\hspace{0.4cm}$0.5762\\
Cl(3a)&$\hspace{0.4cm}$\;0.2912&$\hspace{0.4cm}$\;0.5181&$\hspace{0.4cm}$0.3591\\
Cl(3b)&$\hspace{0.4cm}$\;0.2916&$\hspace{0.4cm}$\;0.9823&$\hspace{0.4cm}$0.3595
\end{tabular}

\subsection{{\afmggau}}
\begin{tabular}{llll}
$\hspace{1.0cm}$&$\hspace{0.8cm}x\hspace{0.8cm}$&$\hspace{0.8cm}y\hspace{0.8cm}$&$\hspace{0.8cm}z\hspace{0.8cm}$\\
Cs(1)&$\hspace{0.4cm}$\;0.1321 &$\hspace{0.4cm}$\;0.2501 &$\hspace{0.4cm}$0.1029\\
Cs(2)&$\hspace{0.4cm}$\;0.9946 &$\hspace{0.4cm}$\;0.7501 &$\hspace{0.4cm}$0.3302\\
Cu&$\hspace{0.4cm}$\;0.2319&$\hspace{0.4cm}$\;0.2501 &$\hspace{0.4cm}$0.4175\\
Cl(1)&$\hspace{0.4cm}$\;0.0037 &$\hspace{0.4cm}$\;0.2499 &$\hspace{0.4cm}$0.3790\\
Cl(2)&$\hspace{0.4cm}$\;0.3446&$\hspace{0.4cm}$\;0.2501 &$\hspace{0.4cm}$0.5777\\
Cl(3a)&$\hspace{0.4cm}$\;0.2959 &$\hspace{0.4cm}$\;0.5151 &$\hspace{0.4cm}$0.3558\\
Cl(3b)&$\hspace{0.4cm}$\;0.2959 &$\hspace{0.4cm}$\;0.9851 &$\hspace{0.4cm}$0.3557
\end{tabular}

\subsection{Experimentally determined structure}
\begin{tabular}{llll}
$\hspace{1.0cm}$&$\hspace{0.8cm}x\hspace{0.8cm}$&$\hspace{0.8cm}y\hspace{0.8cm}$&$\hspace{0.8cm}z\hspace{0.8cm}$\\
Cs(1)&$\hspace{0.4cm}$\;0.1340 &$\hspace{0.4cm}$\;0.25 &$\hspace{0.4cm}$0.1031\\
Cs(2)&$\hspace{0.4cm}$\;0.9433 &$\hspace{0.4cm}$\;0.75 &$\hspace{0.4cm}$0.3252\\
Cu&$\hspace{0.4cm}$\;0.2302&$\hspace{0.4cm}$\;0.25 &$\hspace{0.4cm}$0.4182\\
Cl(1)&$\hspace{0.4cm}$\;0.0050 &$\hspace{0.4cm}$\;0.25 &$\hspace{0.4cm}$0.3820\\
Cl(2)&$\hspace{0.4cm}$\;0.3433&$\hspace{0.4cm}$\;0.25 &$\hspace{0.4cm}$0.5739\\
Cl(3)&$\hspace{0.4cm}$\;0.2936&$\hspace{0.4cm}$\;0.9881&$\hspace{0.4cm}$0.3550
\end{tabular}

\vspace{0.5cm}
In the case of Cs$_2$CuBr$_4$, the lattice constants are
$a=10.195$~\AA, $b=7.965$~\AA, $c=12.936$~\AA\; and the space group is
{\it Pnma}.

\subsection{{\nmlda}}
\begin{tabular}{llll}
$\hspace{1.0cm}$&$\hspace{0.8cm}x\hspace{0.8cm}$&$\hspace{0.8cm}y\hspace{0.8cm}$&$\hspace{0.8cm}z\hspace{0.8cm}$\\
Cs(1)&$\hspace{0.4cm}$\;0.1244&$\hspace{0.4cm}$\;0.25 &$\hspace{0.4cm}$\;0.1030\\
Cs(2)&$\hspace{0.4cm}$\;0.0142 &$\hspace{0.4cm}$\;0.25 &$\hspace{0.4cm}$\;0.6638\\
Cu&$\hspace{0.4cm}$\;0.2345&$\hspace{0.4cm}$\;0.25 &$\hspace{0.4cm}$\;0.4159\\
Br(1)&$\hspace{0.4cm}$\;0.0090&$\hspace{0.4cm}$\;0.25 &$\hspace{0.4cm}$\;0.3715\\
Br(2)&$\hspace{0.4cm}$\;0.3497 &$\hspace{0.4cm}$\;0.25 &$\hspace{0.4cm}$\;0.5751\\
Br(3)&$\hspace{0.4cm}$\;0.2882&$\hspace{0.4cm}$\;0.5267&$\hspace{0.4cm}$\;0.3649
\end{tabular}

\subsection{{\nmgga}}
\begin{tabular}{llll}
$\hspace{1.0cm}$&$\hspace{0.8cm}x\hspace{0.8cm}$&$\hspace{0.8cm}y\hspace{0.8cm}$&$\hspace{0.8cm}z\hspace{0.8cm}$\\
Cs(1)&$\hspace{0.4cm}$\;0.1272&$\hspace{0.4cm}$\;0.25 &$\hspace{0.4cm}$\;0.1072\\
Cs(2)&$\hspace{0.4cm}$\;0.0117 &$\hspace{0.4cm}$\;0.25 &$\hspace{0.4cm}$\;0.6619\\
Cu&$\hspace{0.4cm}$\;0.2322&$\hspace{0.4cm}$\;0.25 &$\hspace{0.4cm}$\;0.4168\\
Br(1)&$\hspace{0.4cm}$\;0.0001&$\hspace{0.4cm}$\;0.25 &$\hspace{0.4cm}$\;0.3756\\
Br(2)&$\hspace{0.4cm}$\;0.3499 &$\hspace{0.4cm}$\;0.25 &$\hspace{0.4cm}$\;0.5787\\
Br(3)&$\hspace{0.4cm}$\;0.2921&$\hspace{0.4cm}$\;0.5257&$\hspace{0.4cm}$\;0.3603
\end{tabular}

\subsection{{\fmgga}}
\begin{tabular}{llll}
$\hspace{1.0cm}$&$\hspace{0.8cm}x\hspace{0.8cm}$&$\hspace{0.8cm}y\hspace{0.8cm}$&$\hspace{0.8cm}z\hspace{0.8cm}$\\
Cs(1)&$\hspace{0.4cm}$\;0.1260&$\hspace{0.4cm}$\;0.25 &$\hspace{0.4cm}$\;0.1059\\
Cs(2)&$\hspace{0.4cm}$\;0.0060 &$\hspace{0.4cm}$\;0.25 &$\hspace{0.4cm}$\;0.6628\\
Cu&$\hspace{0.4cm}$\;0.2312&$\hspace{0.4cm}$\;0.25 &$\hspace{0.4cm}$\;0.4187\\
Br(1)&$\hspace{0.4cm}$-0.0002&$\hspace{0.4cm}$\;0.25 &$\hspace{0.4cm}$\;0.3773\\
Br(2)&$\hspace{0.4cm}$\;0.3456 &$\hspace{0.4cm}$\;0.25 &$\hspace{0.4cm}$\;0.5818\\
Br(3)&$\hspace{0.4cm}$\;0.2977&$\hspace{0.4cm}$\;0.5197&$\hspace{0.4cm}$\;0.3570
\end{tabular}

\subsection{{\fmggau}}
\begin{tabular}{llll}
$\hspace{1.0cm}$&$\hspace{0.8cm}x\hspace{0.8cm}$&$\hspace{0.8cm}y\hspace{0.8cm}$&$\hspace{0.8cm}z\hspace{0.8cm}$\\
Cs(1)&$\hspace{0.4cm}$\;0.1268&$\hspace{0.4cm}$\;0.25 &$\hspace{0.4cm}$\;0.1044\\
Cs(2)&$\hspace{0.4cm}$\;0.0030 &$\hspace{0.4cm}$\;0.25 &$\hspace{0.4cm}$\;0.6637\\
Cu&$\hspace{0.4cm}$\;0.2326&$\hspace{0.4cm}$\;0.25 &$\hspace{0.4cm}$\;0.4180\\
Br(1)&$\hspace{0.4cm}$\;0.0004&$\hspace{0.4cm}$\;0.25 &$\hspace{0.4cm}$\;0.3789\\
Br(2)&$\hspace{0.4cm}$\;0.3435 &$\hspace{0.4cm}$\;0.25 &$\hspace{0.4cm}$\;0.5825\\
Br(3)&$\hspace{0.4cm}$\;0.3004&$\hspace{0.4cm}$\;0.5177&$\hspace{0.4cm}$\;0.3542
\end{tabular}

\subsection{Experimentally determined structure}
\begin{tabular}{llll}
$\hspace{1.0cm}$&$\hspace{0.8cm}x\hspace{0.8cm}$&$\hspace{0.8cm}y\hspace{0.8cm}$&$\hspace{0.8cm}z\hspace{0.8cm}$\\
Cs(1)&$\hspace{0.4cm}$\;0.1290 &$\hspace{0.4cm}$\;0.25 &$\hspace{0.4cm}$0.1058\\
Cs(2)&$\hspace{0.4cm}$\;0.0049 &$\hspace{0.4cm}$\;0.25 &$\hspace{0.4cm}$0.6694\\
Cu&$\hspace{0.4cm}$\;0.2311&$\hspace{0.4cm}$\;0.25 &$\hspace{0.4cm}$0.4187\\
Br(1)&$\hspace{0.4cm}$\;0.0010&$\hspace{0.4cm}$\;0.25 &$\hspace{0.4cm}$0.3819\\
Br(2)&$\hspace{0.4cm}$\;0.3440 &$\hspace{0.4cm}$\;0.25 &$\hspace{0.4cm}$0.5797\\
Br(3)&$\hspace{0.4cm}$\;0.2960&$\hspace{0.4cm}$\;0.5138&$\hspace{0.4cm}$0.3546
\end{tabular}
\section{Appendix B: Further interaction pathways in Cs$_2$CuCl$_4$ considered in the TB model}
\begin{figure}[tb]
\begin{center}
\subfigure {\includegraphics[width=0.35\textwidth]{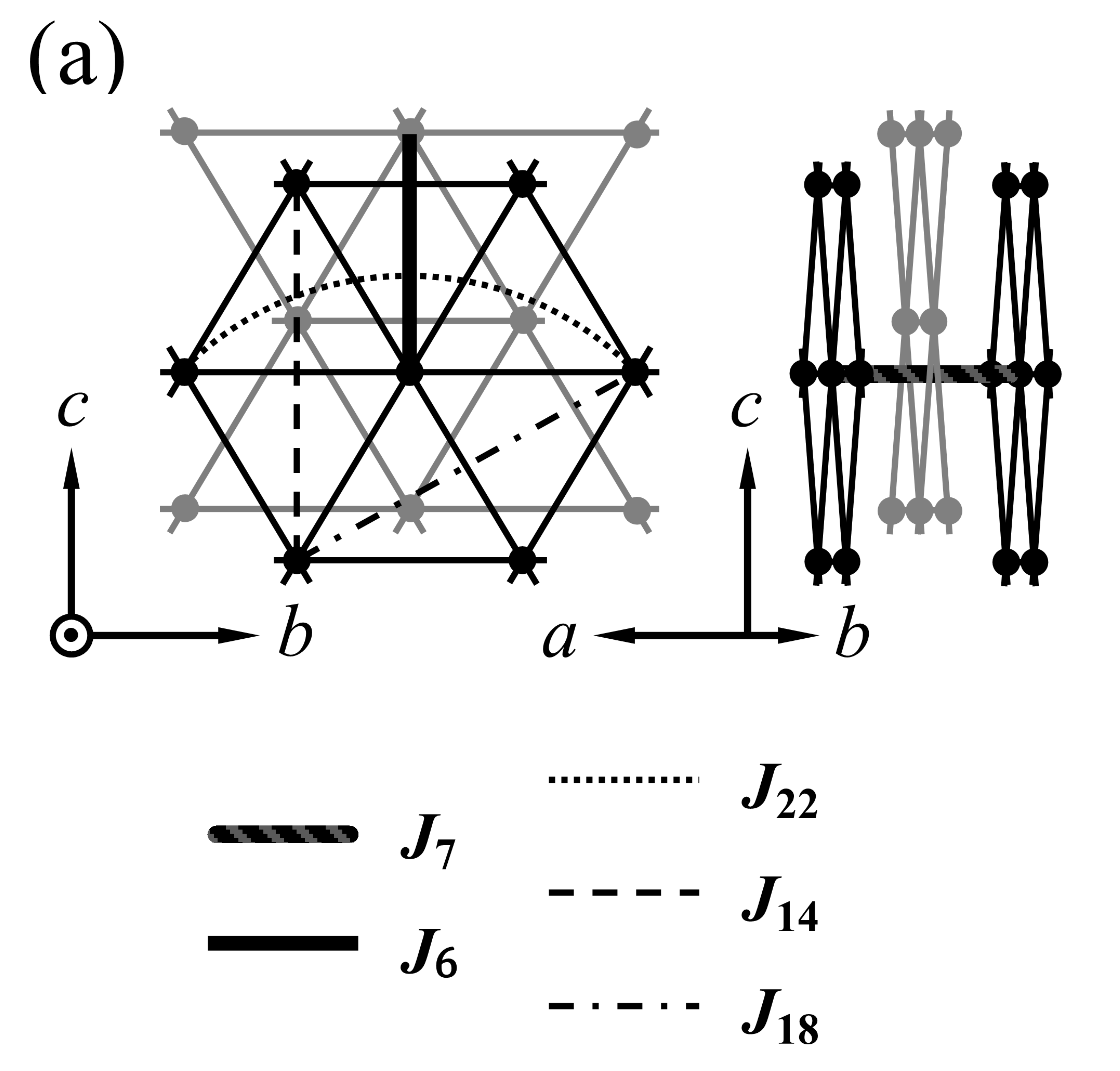}}
\subfigure {\includegraphics[width=0.32\textwidth]{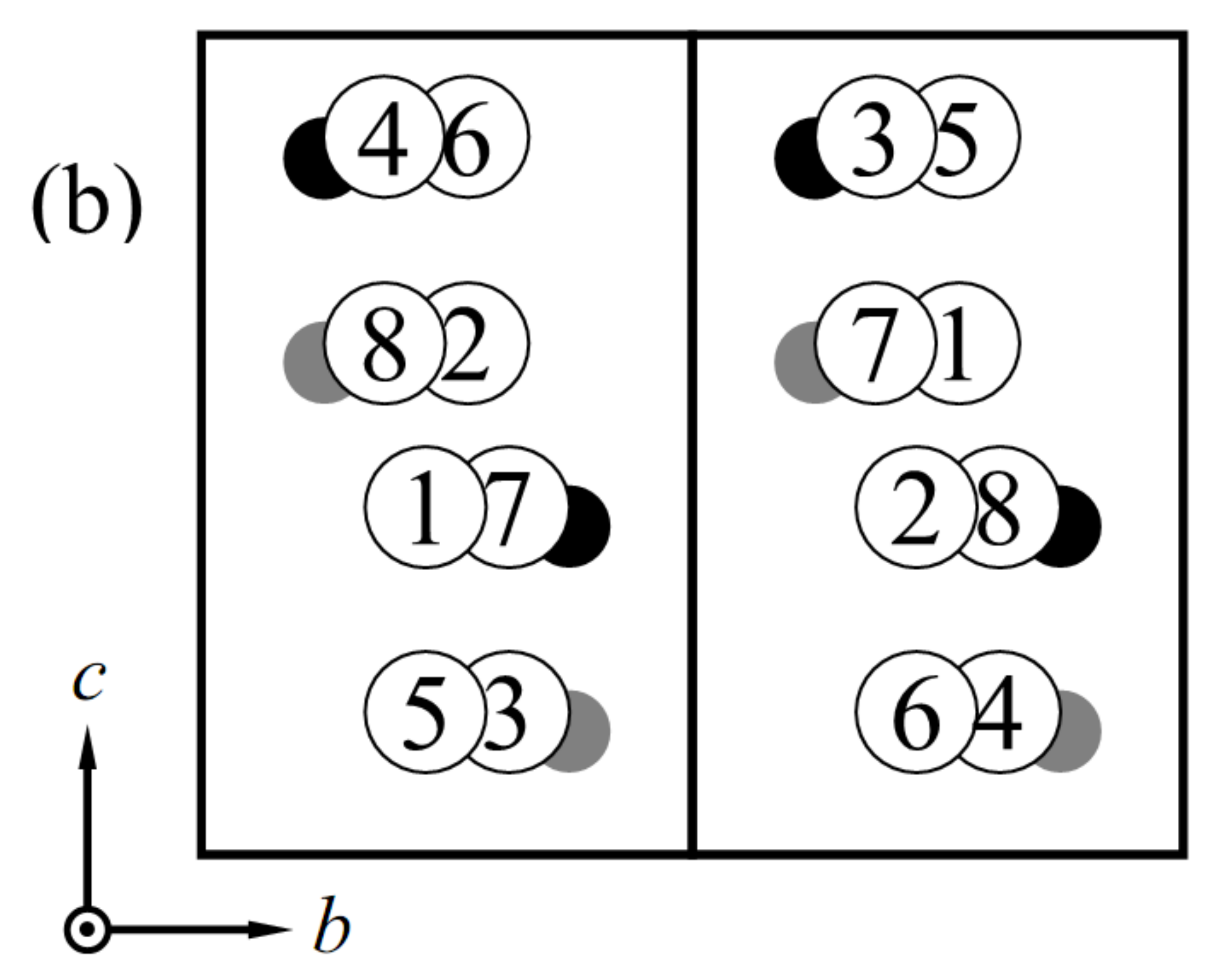}}
\caption{(a) Interaction pathways $J_7$, $J_6$, $J_{14}$, $J_{18}$ and
  $J_{22}$.
(b) Labeling of the Cu atoms in the $2\!\times\!2\!\times\!1$ supercell of
Cs$_2$CuCl$_4$ or   Cs$_2$CuBr$_4$. The black and gray circles denote Cu atoms that
belong to adjacent Cu layers, parallel to the $bc$ plane. Since the
supercell contains two unit cells along the $a$ axis, the black and gray
Cu lattices are doubled, which is not visible in the $bc$ projection. The
two encircled figures, one on top of the other, label the Cu atoms that
have common $y$ and $z$ coordinates, but whose $x$ coordinates differ by the
lattice constant $a$ such that the top figure refers to the Cu atom with
larger $x$ coordinate.
}\label{pathways}
\end{center}
\end{figure}

The interaction pathways for spin exchange coupling constants $J_7$,
$J_6$, $J_{14}$, $J_{18}$ and $J_{22}$ are shown in
Fig.~\ref{pathways}~(a).

\section{Appendix C: Antiferromagnetic spin configurations and corresponding $E^\text{FM}-E^\text{AFM}$ equations for deriving exchange constants in Cs$_2$CuCl$_4$ and Cs$_2$CuBr$_4$}
Given that the Cu atoms in the $2\times2\times1$ supercell of
Cs$_2$CuCl$_4$ or Cs$_2$CuBr$_4$ are labeled as shown in
Fig.~\ref{pathways}~(b), the seven antiferromagnetic (or
ferrimagnetic) spin configurations, considered in order to calculate
exchange couplings $J$, $J'$, $J''_\text{eff}$, $J_1$, $J_3$, $J_7$
and $J_8$ by the total energy difference method, are the following:
\begin{center}
 \begin{tabular}{ccccccccc}
&Cu1 & Cu2 & Cu3 & Cu4 & Cu5 & Cu6 & Cu7 & Cu8 \\
conf. 1: &$\uparrow$&$\downarrow$&$\uparrow$&$\downarrow$&$\uparrow$&$\downarrow$&$\uparrow$&$\downarrow$\\
conf. 2: &$\downarrow$&$\downarrow$&$\uparrow$&$\uparrow$&$\uparrow$&$\uparrow$&$\downarrow$&$\downarrow$\\
conf. 3: &$\uparrow$&$\uparrow$&$\uparrow$&$\uparrow$&$\uparrow$&$\uparrow$&$\downarrow$&$\downarrow$\\
conf. 4: &$\uparrow$&$\uparrow$&$\uparrow$&$\uparrow$&$\downarrow$&$\downarrow$&$\downarrow$&$\downarrow$\\
conf. 5: &$\uparrow$&$\downarrow$&$\downarrow$&$\uparrow$&$\uparrow$&$\downarrow$&$\downarrow$&$\uparrow$\\
conf. 6: &$\downarrow$&$\downarrow$&$\uparrow$&$\uparrow$&$\downarrow$&$\downarrow$&$\uparrow$&$\uparrow$\\
conf. 7: &$\uparrow$&$\downarrow$&$\downarrow$&$\downarrow$&$\downarrow$&$\uparrow$&$\uparrow$&$\uparrow$
\end{tabular}
\end{center}
Each of these spin configurations leads to a corresponding equation:
\begin{equation}
 \begin{array}{rrrrrrrr}
  2J&+2J'&    &+\;\:J_1&+\;\:J_3&+4J_8&     &\displaystyle=E_1/4,\\
    & 2J'&+\;\;J''_\text{eff}&    &    &+2J_8&     &\displaystyle=E_2/8,\\
    & 2J'&+\;\;J''_\text{eff}&+\;\:J_1&    &+2J_8&+\;\:J_7 &\displaystyle=E_3/4,\\
    &    & J''_\text{eff}    &+\;\:J_1&+\;\:J_3&+2J_8&+2J_7 &\displaystyle=E_4/4,\\
  2J&+2J'&+\;\;J''_\text{eff}&+\;\:J_1&+\;\:J_3&+2J_8&+2J_7 &\displaystyle=E_5/4,\\
    & 4J'&+\;\;J''_\text{eff}&+\;\:J_1&+\;\:J_3&+2J_8&+2J_7 &\displaystyle=E_6/4,\\
  4J&+8J'&+6J''_\text{eff}   &+3J_1&+3J_3&+8J_8&+4J_7 &\displaystyle=E_7,
 \end{array}\nonumber
\end{equation}
where $E_i=E^\text{FM}-E^\text{AFM}_i$, $i=1,\ldots,7$, with
$E^\text{FM}$ being the energy of the supercell in the ferromagnetic
configuration of Cu spins and $E^\text{AFM}_i$ being the energy of the
supercell in the antiferromagnetic configuration $i$.

\section{Appendix D: Exchange couplings}
The exchange couplings provided below for the experimental as well as
the {\nmgga}, {\fmgga} and {\fmggau} relaxed Cs$_2$CuCl$_4$ and
Cs$_2$CuBr$_4$ crystal structures are given in meV.
We have marked  in bold face  the parameter values that
should provide a realistic description of the spin models for Cs$_2$CuCl$_4$
and Cs$_2$CuBr$_4$.

\subsection{Cs$_2$CuCl$_4$: experimental structure [Fig.~\ref{J} (b)] }
\begin{tabular}{llll}
\hspace{0.4cm}     & \hspace{0.4cm}$U$=4~eV\hspace{0.4cm} & \hspace{0.4cm}$U$=6~eV\hspace{0.4cm} & \hspace{0.4cm}$U$=8~eV\hspace{0.4cm}\\
& & & \\
\hspace{0.4cm}$J$  &$\hspace{0.4cm}$\;0.2170 &$\hspace{0.4cm}$\;0.1128 &$\hspace{0.4cm}$\;0.0413\\
\hspace{0.4cm}$J'$ &$\hspace{0.4cm}$\;0.0548 &$\hspace{0.4cm}$\;0.0333 &$\hspace{0.4cm}$\;0.0169\\
\hspace{0.4cm}$J''_{\rm eff}$
                   &$\hspace{0.4cm}$\;0.0112 &$\hspace{0.4cm}$\;0.0061 &$\hspace{0.4cm}$\;0.0048\\
\hspace{0.4cm}$J_1$&$\hspace{0.4cm}$\;0.0085 &$\hspace{0.4cm}$-0.0039  &$\hspace{0.4cm}$-0.0046 \\
\hspace{0.4cm}$J_3$&$\hspace{0.4cm}$\;0.0229 &$\hspace{0.4cm}$\;0.0160 &$\hspace{0.4cm}$\;0.0070\\
\hspace{0.4cm}$J_8$&$\hspace{0.4cm}$\;0.0057 &$\hspace{0.4cm}$\;0.0035  &$\hspace{0.4cm}$\;0.0018\\
\hspace{0.4cm}$J_7$&$\hspace{0.4cm}$-0.0152  &$\hspace{0.4cm}$-0.0081  &$\hspace{0.4cm}$-0.0030
\end{tabular}

\subsection{Cs$_2$CuCl$_4$: {\fmggau} relaxed structure [Fig.~\ref{J} (c)] }
\begin{tabular}{llll}
\hspace{0.4cm}     & \hspace{0.4cm}$U$=4~eV\hspace{0.4cm} & \hspace{0.4cm}$U$=6~eV\hspace{0.4cm} & \hspace{0.4cm}$U$=8~eV\hspace{0.4cm}\\
& & & \\
\hspace{0.4cm}$J$  &$\hspace{0.4cm}$\;0.8517 &$\hspace{0.4cm}$\;{\bf 0.3837} &$\hspace{0.4cm}$\;0.1385\\
\hspace{0.4cm}$J'$ &$\hspace{0.4cm}$\;0.3736 &$\hspace{0.4cm}$\;{\bf 0.1788} &$\hspace{0.4cm}$\;0.0779\\
\hspace{0.4cm}$J''_{\rm eff}$
                   &$\hspace{0.4cm}$\;0.0266 &$\hspace{0.4cm}$\;{\bf 0.0137} &$\hspace{0.4cm}$\;0.0089\\
\hspace{0.4cm}$J_1$&$\hspace{0.4cm}$\;0.0210 &$\hspace{0.4cm}$\;{\bf 0.0039} &$\hspace{0.4cm}$\;0.0010\\
\hspace{0.4cm}$J_3$&$\hspace{0.4cm}$\;0.0008 &$\hspace{0.4cm}$\;{\bf 0.0147} &$\hspace{0.4cm}$\;0.0046\\
\hspace{0.4cm}$J_8$&$\hspace{0.4cm}$-0.0039  &$\hspace{0.4cm}${\bf -0.0005}  &$\hspace{0.4cm}$\;0.0011\\
\hspace{0.4cm}$J_7$&$\hspace{0.4cm}$-0.0219  &$\hspace{0.4cm}${\bf -0.0172}  &$\hspace{0.4cm}$-0.0065
\end{tabular}

\subsection{Cs$_2$CuCl$_4$: {\fmgga} relaxed structure [Fig.~\ref{J} (d)] }
\begin{tabular}{llll}
\hspace{0.4cm}     & \hspace{0.4cm}$U$=4~eV\hspace{0.4cm} & \hspace{0.4cm}$U$=6~eV\hspace{0.4cm} & \hspace{0.4cm}$U$=8~eV\hspace{0.4cm}\\
& & & \\
\hspace{0.4cm}$J$  &$\hspace{0.4cm}$\;1.2632 &$\hspace{0.4cm}$\;0.5679 &$\hspace{0.4cm}$\;0.2095\\
\hspace{0.4cm}$J'$ &$\hspace{0.4cm}$\;0.3207 &$\hspace{0.4cm}$\;0.1556 &$\hspace{0.4cm}$\;0.0702\\
\hspace{0.4cm}$J''_{\rm eff}$
                   &$\hspace{0.4cm}$\;0.0224 &$\hspace{0.4cm}$\;0.0175 &$\hspace{0.4cm}$\;0.0118\\
\hspace{0.4cm}$J_1$&$\hspace{0.4cm}$\;0.0308 &$\hspace{0.4cm}$\;0.0127 &$\hspace{0.4cm}$\;0.0031\\
\hspace{0.4cm}$J_3$&$\hspace{0.4cm}$\;0.0435 &$\hspace{0.4cm}$\;0.0302 &$\hspace{0.4cm}$\;0.0108\\
\hspace{0.4cm}$J_8$&$\hspace{0.4cm}$-0.0043  &$\hspace{0.4cm}$-0.0010  &$\hspace{0.4cm}$\;0.0011\\
\hspace{0.4cm}$J_7$&$\hspace{0.4cm}$-0.0222  &$\hspace{0.4cm}$-0.0172  &$\hspace{0.4cm}$-0.0067
\end{tabular}

\subsection{Cs$_2$CuCl$_4$: {\nmgga} relaxed structure [Fig.~\ref{J} (e)] }
\begin{tabular}{llll}
\hspace{0.4cm}     & \hspace{0.4cm}$U$=4~eV\hspace{0.4cm} & \hspace{0.4cm}$U$=6~eV\hspace{0.4cm} & \hspace{0.4cm}$U$=8~eV\hspace{0.4cm}\\
& & & \\
\hspace{0.4cm}$J$  &$\hspace{0.4cm}$\;2.1036 &$\hspace{0.4cm}$\;0.9321 &$\hspace{0.4cm}$\;0.3440\\
\hspace{0.4cm}$J'$ &$\hspace{0.4cm}$\;0.2873 &$\hspace{0.4cm}$\;0.1418 &$\hspace{0.4cm}$\;0.0656\\
\hspace{0.4cm}$J''_{\rm eff}$
                   &$\hspace{0.4cm}$\;0.0076 &$\hspace{0.4cm}$\;0.0089 &$\hspace{0.4cm}$\;0.0112\\
\hspace{0.4cm}$J_1$&$\hspace{0.4cm}$\;0.0424 &$\hspace{0.4cm}$\;0.0150 &$\hspace{0.4cm}$\;0.0061\\
\hspace{0.4cm}$J_3$&$\hspace{0.4cm}$\;0.0919 &$\hspace{0.4cm}$\;0.0468 &$\hspace{0.4cm}$\;0.0185\\
\hspace{0.4cm}$J_8$&$\hspace{0.4cm}$-0.0047  &$\hspace{0.4cm}$\;0.0006 &$\hspace{0.4cm}$\;0.0010\\
\hspace{0.4cm}$J_7$&$\hspace{0.4cm}$-0.0247  &$\hspace{0.4cm}$-0.0138  &$\hspace{0.4cm}$-0.0065
\end{tabular}

\subsection{Cs$_2$CuBr$_4$: experimental structure [Fig.~\ref{J} (f)] }
\begin{tabular}{llll}
\hspace{0.4cm}     & \hspace{0.4cm}$U$=4~eV\hspace{0.4cm} & \hspace{0.4cm}$U$=6~eV\hspace{0.4cm} & \hspace{0.4cm}$U$=8~eV\hspace{0.4cm}\\
& & & \\
\hspace{0.4cm}$J$  &$\hspace{0.4cm}$\;1.6597 &$\hspace{0.4cm}$\;0.7911 &$\hspace{0.4cm}$\;0.2756\\
\hspace{0.4cm}$J'$ &$\hspace{0.4cm}$\;1.0427 &$\hspace{0.4cm}$\;0.5064 &$\hspace{0.4cm}$\;0.2030\\
\hspace{0.4cm}$J''_{\rm eff}$
                   &$\hspace{0.4cm}$\;0.0868 &$\hspace{0.4cm}$\;0.0336 &$\hspace{0.4cm}$\;0.0127\\
\hspace{0.4cm}$J_1$&$\hspace{0.4cm}$\;0.1525 &$\hspace{0.4cm}$\;0.0516 &$\hspace{0.4cm}$\;0.0143\\
\hspace{0.4cm}$J_3$&$\hspace{0.4cm}$\;0.0675 &$\hspace{0.4cm}$\;0.0599 &$\hspace{0.4cm}$\;0.0273\\
\hspace{0.4cm}$J_8$&$\hspace{0.4cm}$-0.0147  &$\hspace{0.4cm}$-0.0002  &$\hspace{0.4cm}$\;0.0019\\
\hspace{0.4cm}$J_7$&$\hspace{0.4cm}$-0.0266  &$\hspace{0.4cm}$-0.0116  &$\hspace{0.4cm}$-0.0040
\end{tabular}

\subsection{Cs$_2$CuBr$_4$: {\fmggau} relaxed structure [Fig.~\ref{J} (g)] }
\begin{tabular}{llll}
\hspace{0.4cm}     & \hspace{0.4cm}$U$=4~eV\hspace{0.4cm} & \hspace{0.4cm}$U$=6~eV\hspace{0.4cm} & \hspace{0.4cm}$U$=8~eV\hspace{0.4cm}\\
& & & \\
\hspace{0.4cm}$J$  &$\hspace{0.4cm}$\;2.0358 &$\hspace{0.4cm}$\;{\bf 0.9282} &$\hspace{0.4cm}$\;0.3019 \\
\hspace{0.4cm}$J'$ &$\hspace{0.4cm}$\;1.0223 &$\hspace{0.4cm}$\;{\bf 0.4821} &$\hspace{0.4cm}$\;0.1815 \\
\hspace{0.4cm}$J''_{\rm eff}$
                   &$\hspace{0.4cm}$\;0.0589 &$\hspace{0.4cm}$\;{\bf 0.0238} &$\hspace{0.4cm}$\;0.0105\\
\hspace{0.4cm}$J_1$&$\hspace{0.4cm}$\;0.1268 &$\hspace{0.4cm}$\;{\bf 0.0436} &$\hspace{0.4cm}$\;0.0119\\
\hspace{0.4cm}$J_3$&$\hspace{0.4cm}$\;0.0685 &$\hspace{0.4cm}$\;{\bf 0.0694} &$\hspace{0.4cm}$\;0.0307\\
\hspace{0.4cm}$J_8$&$\hspace{0.4cm}$-0.0142  &$\hspace{0.4cm}${\bf -0.0011}  &$\hspace{0.4cm}$\;0.0012\\
\hspace{0.4cm}$J_7$&$\hspace{0.4cm}$-0.0368  &$\hspace{0.4cm}${\bf -0.0204}  &$\hspace{0.4cm}$-0.0086
\end{tabular}

\subsection{Cs$_2$CuBr$_4$: {\fmgga} relaxed structure [Fig.~\ref{J} (h)] }
\begin{tabular}{llll}
\hspace{0.4cm}     & \hspace{0.4cm}$U$=4~eV\hspace{0.4cm} & \hspace{0.4cm}$U$=6~eV\hspace{0.4cm} & \hspace{0.4cm}$U$=8~eV\hspace{0.4cm}\\
& & & \\
\hspace{0.4cm}$J$  &$\hspace{0.4cm}$\;2.7229 &$\hspace{0.4cm}$\;1.2325 &$\hspace{0.4cm}$\;0.4076 \\
\hspace{0.4cm}$J'$ &$\hspace{0.4cm}$\;0.9234 &$\hspace{0.4cm}$\;0.4382 &$\hspace{0.4cm}$\;0.1682 \\
\hspace{0.4cm}$J''_{\rm eff}$
                   &$\hspace{0.4cm}$\;0.0554 &$\hspace{0.4cm}$\;0.0219 &$\hspace{0.4cm}$\;0.0122\\
\hspace{0.4cm}$J_1$&$\hspace{0.4cm}$\;0.1467 &$\hspace{0.4cm}$\;0.0492 &$\hspace{0.4cm}$\;0.0166\\
\hspace{0.4cm}$J_3$&$\hspace{0.4cm}$\;0.1710 &$\hspace{0.4cm}$\;0.1101 &$\hspace{0.4cm}$\;0.0468\\
\hspace{0.4cm}$J_8$&$\hspace{0.4cm}$-0.0158  &$\hspace{0.4cm}$-0.0005  &$\hspace{0.4cm}$\;0.0004\\
\hspace{0.4cm}$J_7$&$\hspace{0.4cm}$-0.0394  &$\hspace{0.4cm}$-0.0183  &$\hspace{0.4cm}$-0.0089
\end{tabular}

\subsection{Cs$_2$CuBr$_4$: {\nmgga} relaxed structure [Fig.~\ref{J} (i)] }
\begin{tabular}{llll}
\hspace{0.4cm}     & \hspace{0.4cm}$U$=4~eV\hspace{0.4cm} & \hspace{0.4cm}$U$=6~eV\hspace{0.4cm} & \hspace{0.4cm}$U$=8~eV\hspace{0.4cm}\\
& & & \\
\hspace{0.4cm}$J$  &$\hspace{0.4cm}$\;5.1309 &$\hspace{0.4cm}$\;2.2669 &$\hspace{0.4cm}$\;0.7423 \\
\hspace{0.4cm}$J'$ &$\hspace{0.4cm}$\;0.8025 &$\hspace{0.4cm}$\;0.3872 &$\hspace{0.4cm}$\;0.1540 \\
\hspace{0.4cm}$J''_{\rm eff}$
                   &$\hspace{0.4cm}$\;0.0356 &$\hspace{0.4cm}$\;0.0129 &$\hspace{0.4cm}$\;0.0115\\
\hspace{0.4cm}$J_1$&$\hspace{0.4cm}$\;0.1980 &$\hspace{0.4cm}$\;0.0696 &$\hspace{0.4cm}$\;0.0246\\
\hspace{0.4cm}$J_3$&$\hspace{0.4cm}$\;0.3723 &$\hspace{0.4cm}$\;0.1907 &$\hspace{0.4cm}$\;0.0741\\
\hspace{0.4cm}$J_8$&$\hspace{0.4cm}$-0.0196  &$\hspace{0.4cm}$-0.0015  &$\hspace{0.4cm}$\;0.0011\\
\hspace{0.4cm}$J_7$&$\hspace{0.4cm}$-0.0521  &$\hspace{0.4cm}$-0.0190  &$\hspace{0.4cm}$-0.0083
\end{tabular}


\begin{thebibliography}{99}
\bibitem{remark}  Note however that the definition
of local site moments is not unique and may vary to some degree between
different band structure codes.
\end{thebibliography}
\end{document}